\documentclass[11pt,preprint]{aastex}

\shorttitle{Angular Diameter of HR 8799}
\shortauthors{Baines et al.}

\usepackage{amsmath}
\begin{document}

%\journalinfo{Astrophysical Journal, accepted}
%\submitted{Astrophysical Journal, accepted}

\title{The CHARA Array Angular Diameter of HR 8799 Favors Planetary Masses for Its Imaged Companions }

\author{Ellyn K. Baines}
\affil{Remote Sensing Division, Naval Research Laboratory, 4555 Overlook Avenue SW, \\ Washington, DC 20375}
\email{ellyn.baines@nrl.navy.mil}

\author{Russel J. White}
\affil{Center for High Angular Resolution Astronomy, Georgia State University, P.O. Box 3969, \\ Atlanta, GA 30302-3969}

\author{Daniel Huber}
\affil{NASA Ames Research Center, Moffett Field, CA 94035}

\author{Jeremy Jones, Tabetha Boyajian, Harold A. McAlister, \\ Theo A. ten Brummelaar, Nils H. Turner, Judit Sturmann, Laszlo Sturmann, P. J. Goldfinger, Christopher D. Farrington, Adric R. Riedel}
\affil{Center for High Angular Resolution Astronomy, Georgia State University, P.O. Box 3969, \\ Atlanta, GA 30302-3969}

\author{Michael Ireland}
\affil{Department of Physics \& Astronomy, Macquarie University, NSW 2109, New South Wales, Australia} 

\author{Kaspar von Braun}
\affil{NASA Exoplanet Science Institute, California Institute of Technology, 770 S. Wilson Ave, MS 100-22, Pasadena, CA 91125-2200}

\author{Stephen T. Ridgway}
\affil{Kitt Peak National Observatory, National Optical Astronomy Observatory, P.O. Box 26732, Tucson, AZ 85726-6732} 

\begin{abstract}

HR 8799 is an hF0 mA5 $\gamma$ Doradus, $\lambda$ Bootis, Vega-type star best known for hosting four directly imaged candidate planetary companions. Using the CHARA Array interferometer, we measure HR 8799's limb-darkened angular diameter to be 0.342$\pm$0.008 mas; this is the smallest interferometrically measured stellar diameter to date, with an error of only 2$\%$. By combining our measurement with the star's parallax and photometry from the literature, we greatly improve upon previous estimates of its fundamental parameters, including stellar radius (1.44$\pm$0.06 $R_\odot$), effective temperature (7193$\pm$87 K, consistent with F0), luminosity (5.05$\pm$0.29 $L_\odot$), and the extent of the habitable zone (1.62 AU to 3.32 AU). These improved stellar properties permit much more precise comparisons with stellar evolutionary models, from which a mass and age can be determined, once the metallicity of the star is known. Considering the observational properties of other $\lambda$ Bootis stars and the indirect evidence for youth of HR 8799, we argue that the internal abundance, and what we refer to as the \textit{effective} abundance, is most likely near-solar. Finally, using the Yonsei-Yale evolutionary models with uniformly scaled solar-like abundances, we estimate HR 8799's mass and age considering two possibilities:  $1.516^{+0.038}_{-0.024}$ $M_{\odot}$ and $33^{+7}_{-13.2}$ Myr if the star is contracting toward the zero age main-sequence or $1.513^{+0.023}_{-0.024}$ $M_{\odot}$ and $90^{+381}_{-50}$ Myr if it is expanding from it. This improved estimate of HR 8799's age with realistic uncertainties provides the best constraints to date on the masses of its orbiting companions, and strongly suggests they are indeed planets. They nevertheless all appear to orbit well outside the habitable zone of this young star. 

\end{abstract}


%%%%%%%%%%%%%%%%%%%%%%%%%%%%% Introduction %%%%%%%%%%%%%%%%%%%%%%%%%%%%%
\section{Introduction}
HR 8799 (HD 218396, HIP 114189) is a $\gamma$ Doradus-type star, a class characterized by pulsations on time-scales longer than the typical $\delta$ Scuti stars, their neighbors on the Hertzsprung-Russell (H-R) diagram \citep{1979PASP...91....5B, 1999MNRAS.303..275Z, 1999PASP..111..840K, 1999MNRAS.309L..19H}. HR 8799 is also a $\lambda$ Bootis-type star, which is a spectroscopically-defined group of non-magnetic, chemically peculiar, Population I A- and early F-type stars that show metal deficiencies in Fe-peak elements while showing solar to slightly over solar abundances of C, N, O, and S \citep{1990ApJ...363..234V, 1998AandA...335..533P, 1999AJ....118.2993G}. In addition, HR 8799 is a ``Vega-like'' star, possessing excess infrared emission longward of $\sim 20$ $\mu$m \citep{1986PASP...98..685S, 2009ApJ...705..314S}, attributed to thermal emission from a debris disk.

Aside from these three distinguishing characteristics, interest in HR 8799 was amplified in 2008 with the discovery of three planet-like companions at projected separations of 24, 38 and 68 AU \citep{2008Sci...322.1348M}. From comparisons of the observed fluxes to the predictions of evolutionary models, Marois et al. estimated companion masses of 5-11~$M_{\rm Jupiter}$ for companion b and 7-13~$M_{\rm Jupiter}$ for companions c and d, each. Because the masses of the companions depend on the age of the system, Marois et al. used four lines of reasoning to estimate the host star's age: (1) comparing the star's galactic space motion to other young stars in the solar neighborhood, (2) placing the star on a color-magnitude diagram, (3) characterizing the typical ages of $\lambda$ Boo and $\gamma$ Dor stars, and (4) the assumption that HR 8799 has a massive debris disk, which is typically only found associated with young stars ($\lesssim 500$ Myr). Based on these characteristics, they constrained an age range between 30 and 160 Myr. A fourth companion was imaged two years later with an estimated mass of 7$^{+3}_{-2}$ $M_{\rm Jupiter}$ if the system is 30 Myr old and 10$\pm 3$ $M_{\rm Jupiter}$ if 60 Myr old \citep{2010Natur.468.1080M}. 

\citet{2012ApJ...754..135M} fit atmospheric and evolution models to the data to determine the masses, radii, temperatures, gravities, and cloud properties for the planets. They found masses of 26 $M_{\rm Jupiter}$ for planet b and 8-11 $M_{\rm Jupiter}$ for planets c and d, though they acknowledge that their model fit for b's mass is not likely to be the true value. They also determined an age range of 360 Myr, 40-100 My, and 30-100 Myr for each planet, respectively.

The ages adopted by \citet{2008Sci...322.1348M, 2010Natur.468.1080M} have become a topic of some debate. The star's age is of vital interest because it is directly linked to the masses of the companions. If the star is young, its companions are brighter and their inferred masses are lower -- i.e., planets -- than if the star is older, in which case its more massive companions -- i.e., brown dwarfs -- have cooled significantly and could be mistaken for planets \citep{2010ApJ...721L.199M}. For example, if HR 8799 is as old as the Hyades \citep[625 Myr,][]{1998AandA...331...81P}, all companions would be more massive than 13.6 $M_{\rm Jupiter}$, and thus more appropriately called brown dwarfs according to popular convention. One example of the effect of presumed ages on masses involved the announcement by \citet{2010ApJ...719..497L} of a directly imaged planetary companion to 1RXS J160929.1--210524 in the Upper Scorpius association, which relied on an assumed cluster age of 5 Myr. Soon afterward, \citet{2012ApJ...746..154P} determined an older age for the cluster: 11$^{+1}_{-2}$ Myr.  In this case, the imaged companion has a mass of $\sim$14 $M_{\rm Jupiter}$, putting it just above the planetary mass limit.

\citet{2010MNRAS.405L..81M} refute the \citet{2008Sci...322.1348M} young age of HR 8799 on all four counts. They point out that (1) space motions of young disk stars are often inconclusive, (2) the uncertain internal metallicity of HR 8799 makes comparisons with evolutionary models and other stellar clusters unreliable, (3) that $\lambda$ Boo and $\gamma$ Dor type stars have a broad range of ages, and (4) that debris disks are ``highly chaotic and unpredictable'', and cannot be used to estimate stellar age. \citet{2010MNRAS.405L..81M} instead determined HR 8799's age by modeling $\gamma$ Dor pulsations detected by \citet{1999MNRAS.303..275Z} and adopting spectroscopic parameters by \citet{1999AandA...352..555A}. Although primarily limited by the unknown inclination of the star's rotation axis, they find a preferred age of between 1.1 and 1.5 Gyr; again, this age implies that the companions to HR 8799 are brown dwarfs.

\citet{2010ApJ...721L.199M} subsequently challenged the claim made by \citet{2010MNRAS.405L..81M}, finding an age of the system of $\sim$100 Myr after considering the limiting case for orbital stability; at these relatively small separations, massive companions would not be dynamically stable on long timescales. They used the double 4:2:1 mean motion resonance configuration presented by other authors \citep[e.g.,][]{2009AandA...503..247R,2009MNRAS.397L..16G,2010ApJ...710.1408F} as a way to ensure dynamical stability by avoiding close encounters. Moro-Mart{\'i}n et al. used evolutionary models to predict ages from 150 to 350 Myr and conclude that the dynamical state of the system favors an age of $\sim$100 Myr, which places the companions just shy of the brown dwarf regime. However, they point out that the average temperature of 7350 K and luminosity of 5 $L_\odot$ from \citet{2006PASJ...58.1023S} and \citet{2010MNRAS.406..566M} implies an age for the star that is very young when compared to other $\lambda$ Boo stars.

In addition to these recent attempts to estimate HR 8799's age, many other efforts have been made to determine its age in the last decade. The resulting values vary from 20 Myr up to 1.5 Gyr. In Table \ref{lit_params}, we summarize both the attempts described above and other efforts. A variety of methods were utilized, including inspecting the star's proper motion, comparing its position on an H-R diagram with isochrones, inspecting the star for infrared excess, asteroseismology, astrometry, spectroscopy, dynamical stability analysis, disk symmetry, direct imaging, and group membership. 

Here we seek to greatly improve our understanding of HR 8799's age by using interferometric observations to directly measure its angular diameter. This value, when combined with the \emph{HIPPARCOS} parallax, yields the star's physical radius. In combination with HR 8799's bolometric flux determined from broad-band photometry, the stellar luminosity and effective temperature are precisely determined. Using these results we provide new age and mass estimates based on comparisons with the stellar evolutionary models, adopting a metallicity most appropriate for $\lambda$ Boo stars and HR 8799 in particular. Specifically, Section 2 describes the interferometric observations and calibrator star selection; Section 3 discusses the visibility measurements and how stellar parameters were calculated, including angular diameter, radius, luminosity, and temperature; Section 4 explores the physical implications of the interferometric observations; and Section 5 summarizes our findings.

%%%%%%%%%%%%%%%%%%%%% Interferometric observations %%%%%%%%%%%%%%%%%
\section{Interferometric Observations}
Observations were obtained using the Center for High Angular Resolution Astronomy (CHARA) Array, a six element optical-infrared interferometer located on Mount Wilson, California \citep{2005ApJ...628..453T}. All observations used the Precision Astronomical Visible Observations (PAVO) beam combiner with a spectral bandpass of $\sim$650-800 nm. For a description of the PAVO instrument and data reduction procedure, see \citet{2008SPIE.7013E..63I}. We observed HR 8799 over seven nights spanning two years with the S2-W2 and S1-E1 telescope pairs with maximum baselines of 177 m and 331 m, respectively.\footnote{The three arms of the CHARA Array are denoted by their cardinal directions: ``S'', ``E'', and ``W'' are south, east, and west, respectively. Each arm bears two telescopes, numbered ``1'' for the telescope farthest from the beam combining laboratory and ``2'' for the telescope closer to the lab. The ``baseline'' is the distance between the telescopes.} 

We interleaved calibrator star and HR 8799 observations so that every target was flanked by calibrator observations made as close in time as possible, which allowed us to convert instrumental target and calibrator visibilities to calibrated visibilities for the target. Calibrators are stars with predicted diameters that are significantly smaller and in close proximity in the sky to the target star. Reliable calibrators were chosen to be single stars with angular diameter estimates $\lesssim$0.3 mas so they were nearly unresolved on the baseline used, which meant uncertainties in each calibrator's diameter did not affect the target's diameter calculation as much as if the calibrator star had a larger angular size.

In order to estimate the reddening of each calibrator star, we obtained the Tycho ($B_{\rm T}-V_{\rm T}$) color from \citet{1997ESASP1200.....P} and converted to ($B-V$) using the table in \citet{2000PASP..112..961B}.
We then compared the observed ($B-V$) value with the list of intrinsic ($B-V$) colors as a function of spectral type given by \citet{Schmidt-Kaler}\footnote{The table was obtained from http://obswww.unige.ch/gcpd/mk01bv.html.} to arrive at an estimate of E($B-V$), and adopted the reddening law described in \citet{1994ApJ...422..158O} to de-redden the observed magnitudes. The photometric angular diameters were then determined using the relationship between the ($V-K$) color and log $\theta_{\rm LD}$ from \citet{2004AandA...426..297K}. The error in the diameters is due to the relative calibration error stated in the \citet{2004AandA...426..297K} paper as well as errors in photometry measurements. Table \ref{cals} lists the input photometry and resulting photometrically estimated angular diameters of the calibrator stars.

%%%%%%%%%%%%%%%%% Angular diameter determinations %%%%%%%%%%%%%%%%%%
\section{Results}
\subsection{Angular Diameter Measurement}
The observed quantity of an interferometer is fringe contrast or ``visibility'', more formally defined as the correlation of two wave-fronts whose amplitude is the visibility squared ($V^2$), which is fit to a model of a uniformly-illuminated disk (UD) that represents the face of the star. Diameter fits to $V^2$ were based upon the UD approximation given by $V^2 = ([2 J_1(x)] / x)^2$, where $J_1$ is the first-order Bessel function and $x = \pi B \theta_{\rm UD} \lambda^{-1}$, where $B$ is the projected baseline at the star's position, $\theta_{\rm UD}$ is the apparent UD angular diameter of the star, and $\lambda$ is the effective wavelength of the observation \citep{1992ARAandA..30..457S}. A more realistic model of a star's disk involves limb-darkening (LD), and the relationship incorporating the linear LD coefficient $\mu_{\lambda}$ \citep{1974MNRAS.167..475H} is:
\begin{equation}
V^2 = \left( {1-\mu_\lambda \over 2} + {\mu_\lambda \over 3} \right)^{-2}
\times
\left[(1-\mu_\lambda) {J_1(\rm x) \over \rm x} + \mu_\lambda {\left( \frac{\pi}{2} \right)^{1/2} \frac{J_{3/2}(\rm x)}{\rm x^{3/2}}} \right]^{2} .
\end{equation}
Table \ref{calib_visy} lists the date of observation, the calibrator used, $\lambda$/$B$, the calibrated visibilities ($V^2$), and errors in $V^2$ ($\sigma V^2$).

\begin{figure}[h]
\includegraphics[width=1.0\textwidth]{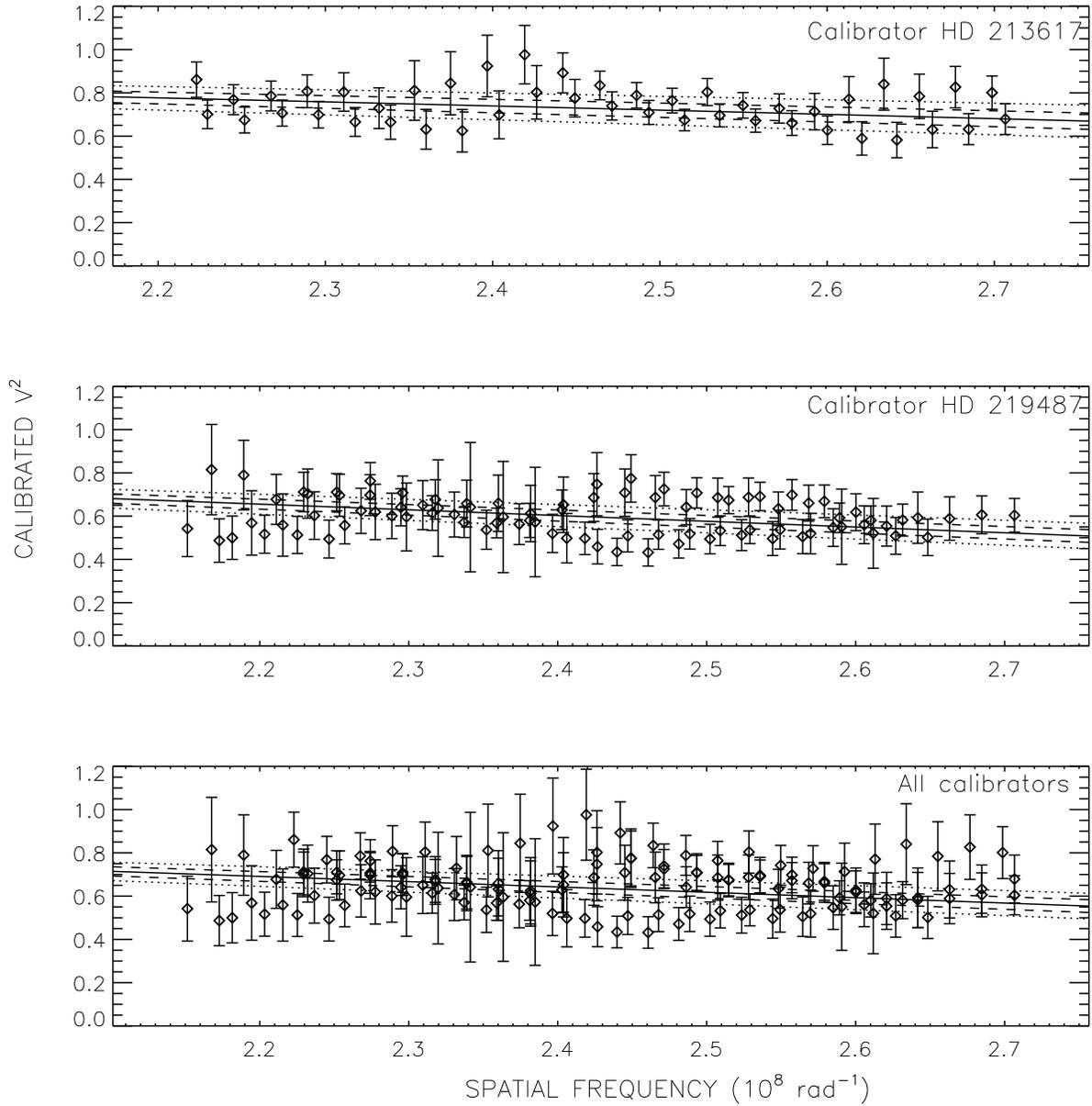}
\caption{$\theta_{\rm LD}$ fit for HR 8799 on 2010 Aug 25. The diamonds and vertical lines are the measured visibilities and their associated errors, the solid line is the best-fit LD diameter, the dashed line is the 1-$\sigma$ error, and the dotted line is the 2-$\sigma$ error.}
  \label{HD218396_072510}
\end{figure}

\begin{figure}[h]
\includegraphics[width=1.0\textwidth]{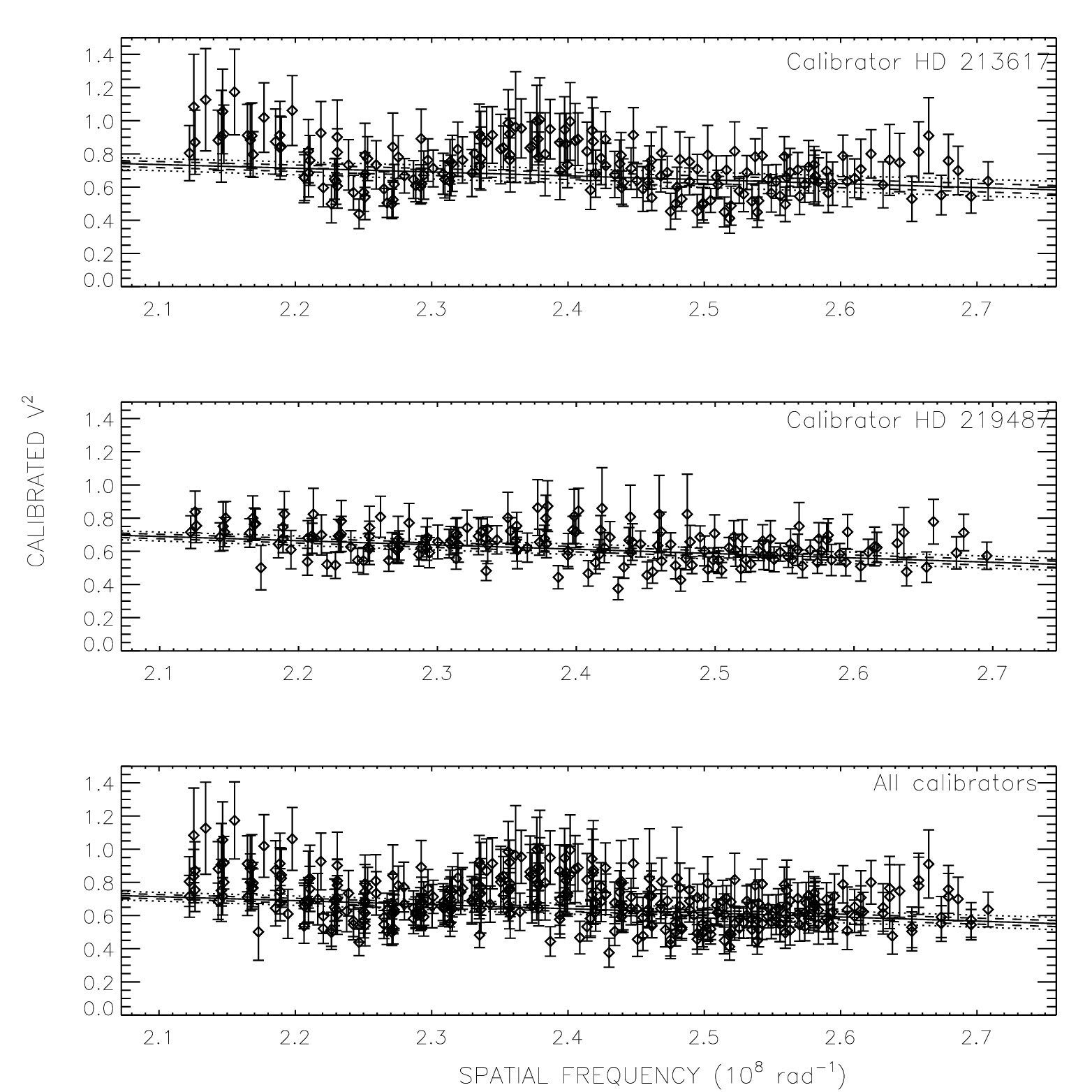}
\caption{$\theta_{\rm LD}$ fit for HR 8799 on 2010 Sep 07. The symbols are the same as in Figure \ref{HD218396_072510}.}
  \label{HD218396_090710}
\end{figure}

\begin{figure}[h]
\includegraphics[width=1.0\textwidth]{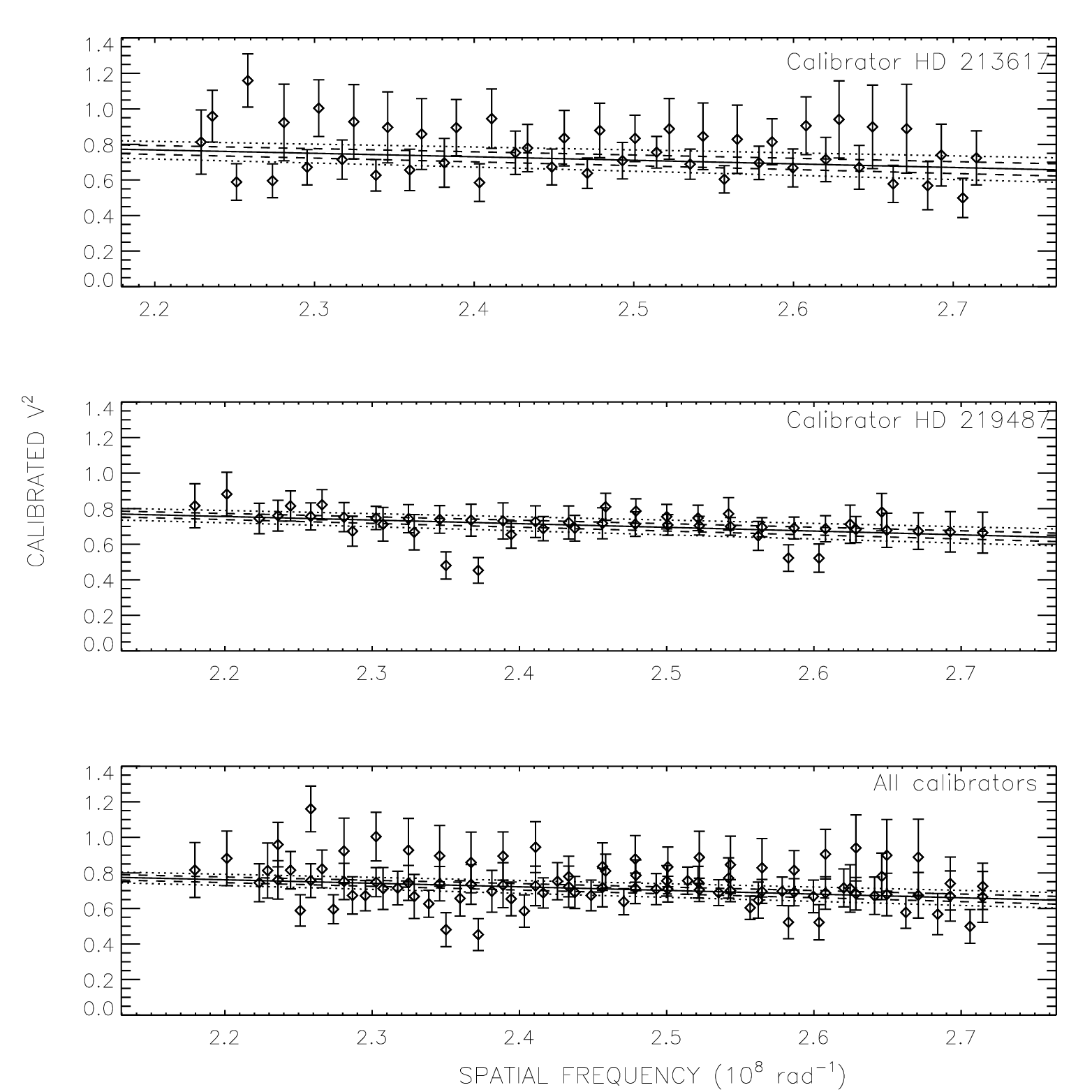}
\caption{$\theta_{\rm LD}$ fit for HR 8799 on 2010 Sep 08. The symbols are the same as in Figure \ref{HD218396_072510}.}
  \label{HD218396_090810}
\end{figure}

\begin{figure}[h]
\includegraphics[width=1.0\textwidth]{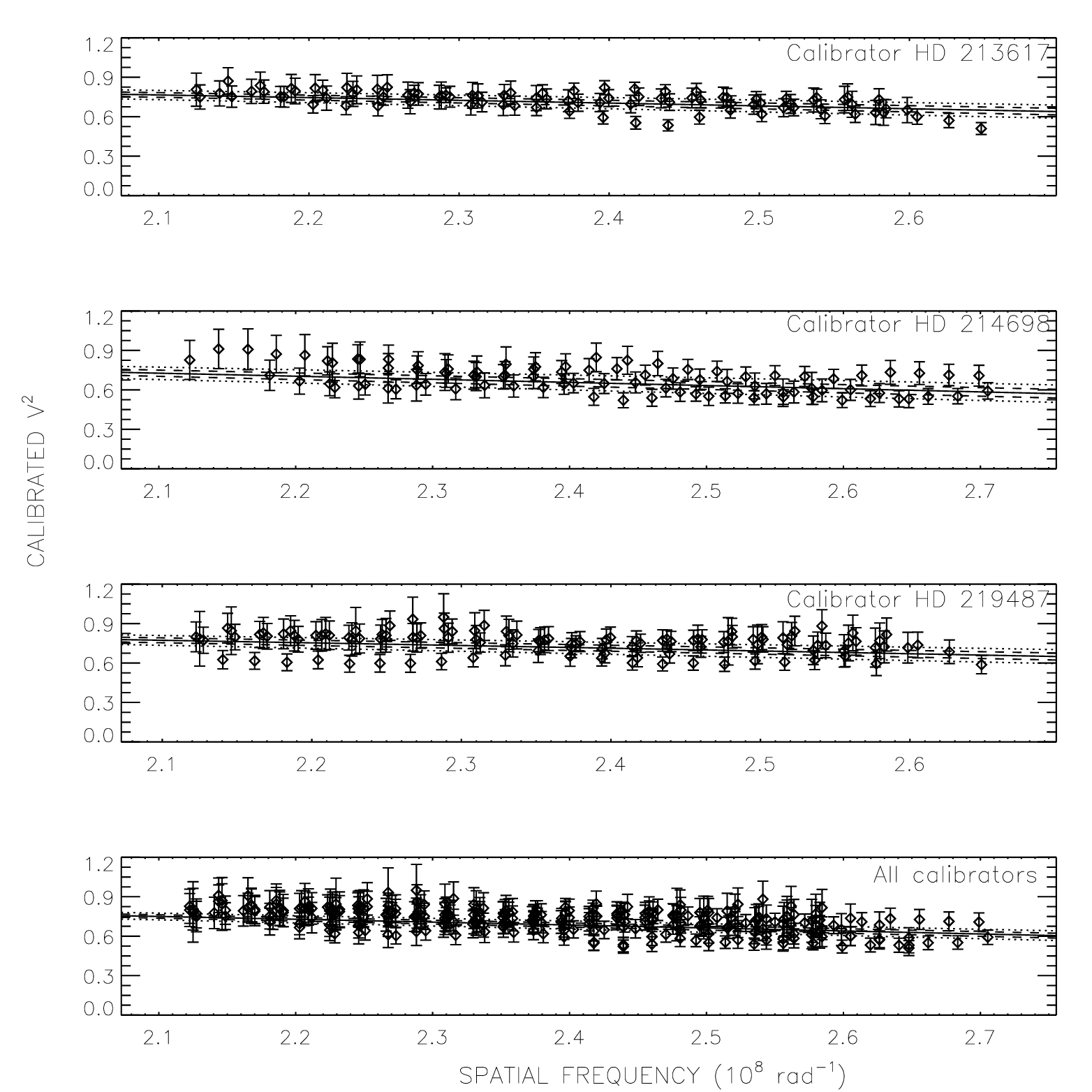}
\caption{$\theta_{\rm LD}$ fit for HR 8799 on 2011 Oct 20. The symbols are the same as in Figure \ref{HD218396_072510}.}
  \label{HD218396_082011}
\end{figure}

\begin{figure}[h]
\includegraphics[width=1.0\textwidth]{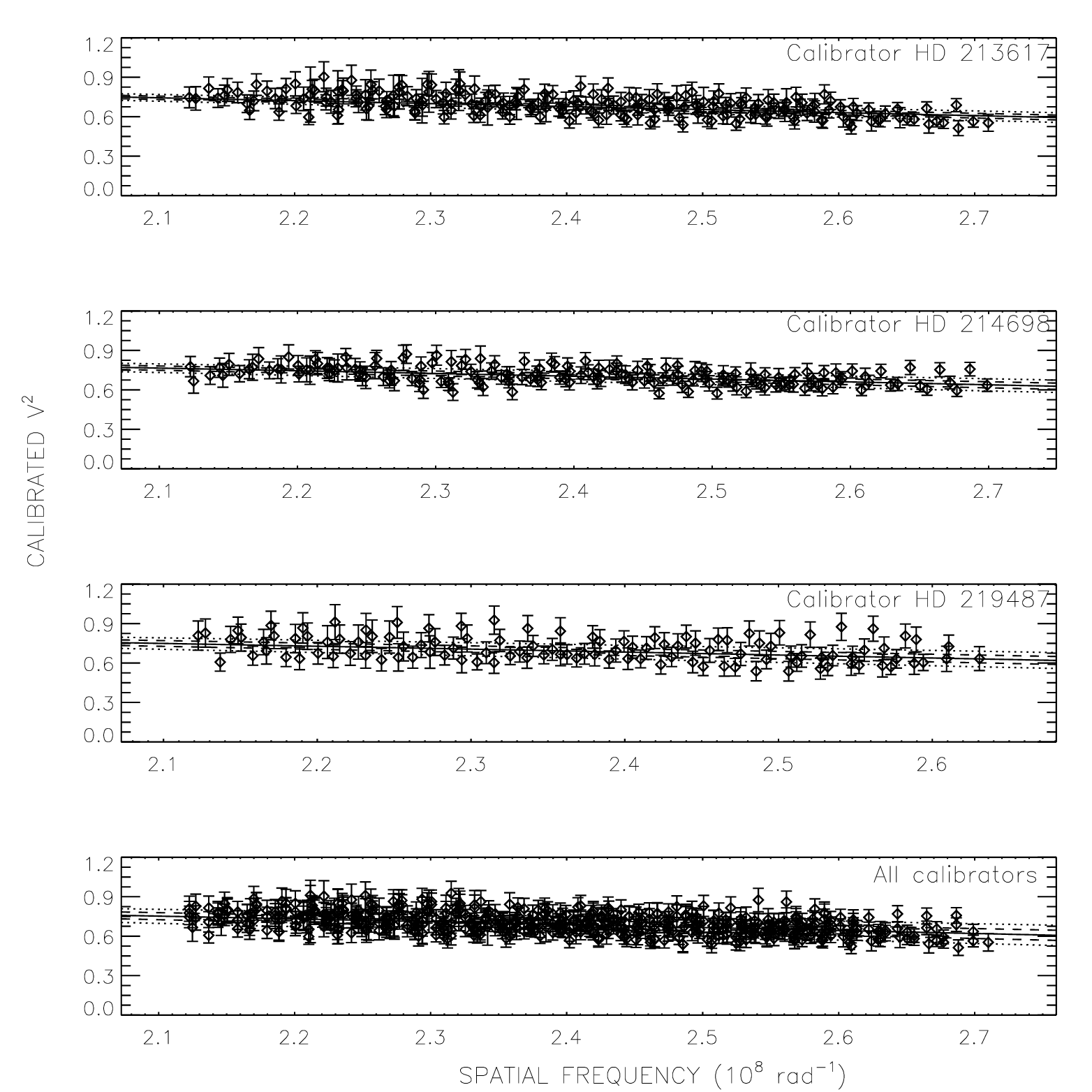}
\caption{$\theta_{\rm LD}$ fit for HR 8799 on 2011 Oct 21. The symbols are the same as in Figure \ref{HD218396_072510}.}
  \label{HD218396_082111}
\end{figure}

\begin{figure}[h]
\includegraphics[width=1.0\textwidth]{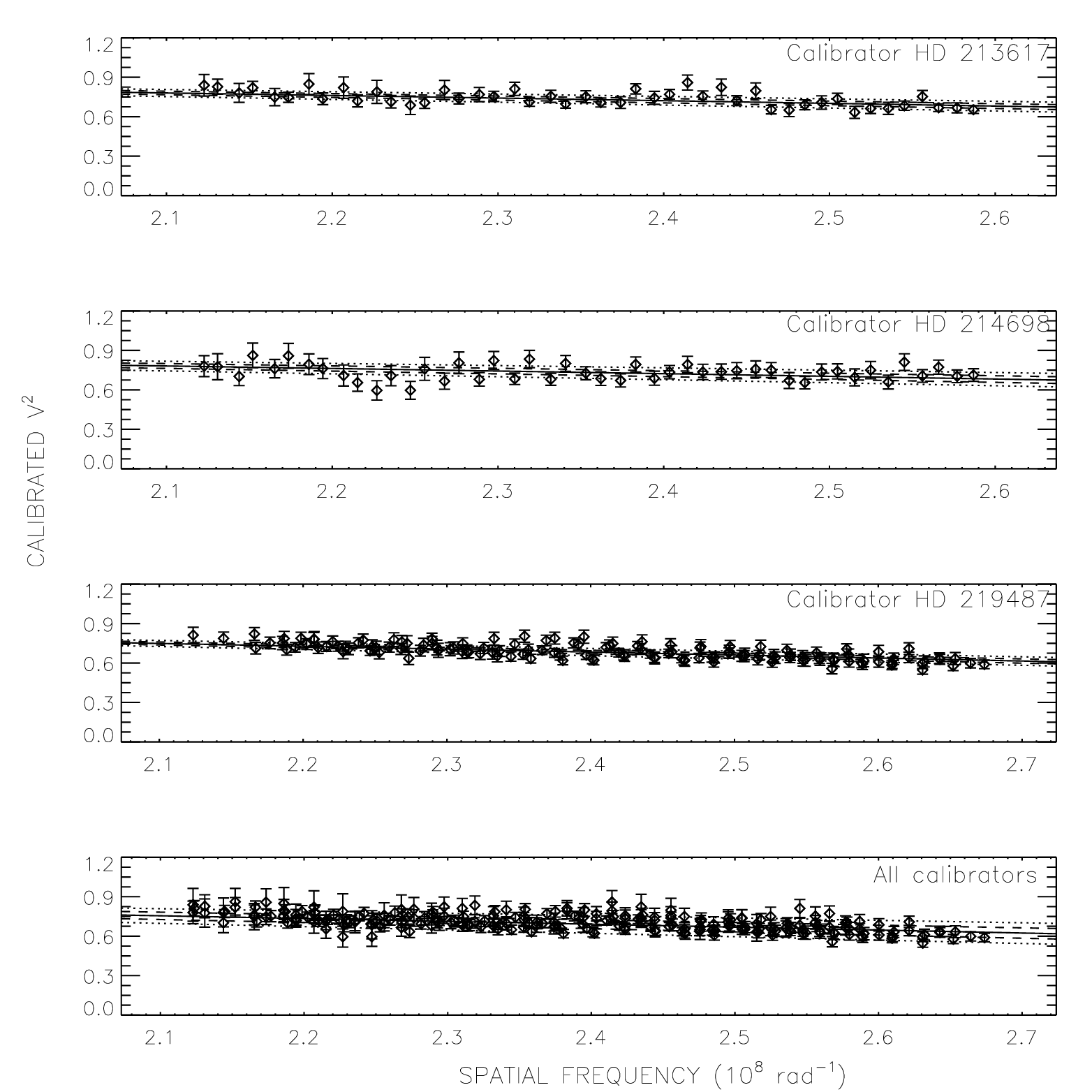}
\caption{$\theta_{\rm LD}$ fit for HR 8799 on 2011 Oct 22. The symbols are the same as in Figure \ref{HD218396_072510}.}
  \label{HD218396_082211}
\end{figure}

\begin{figure}[h]
\includegraphics[width=1.0\textwidth]{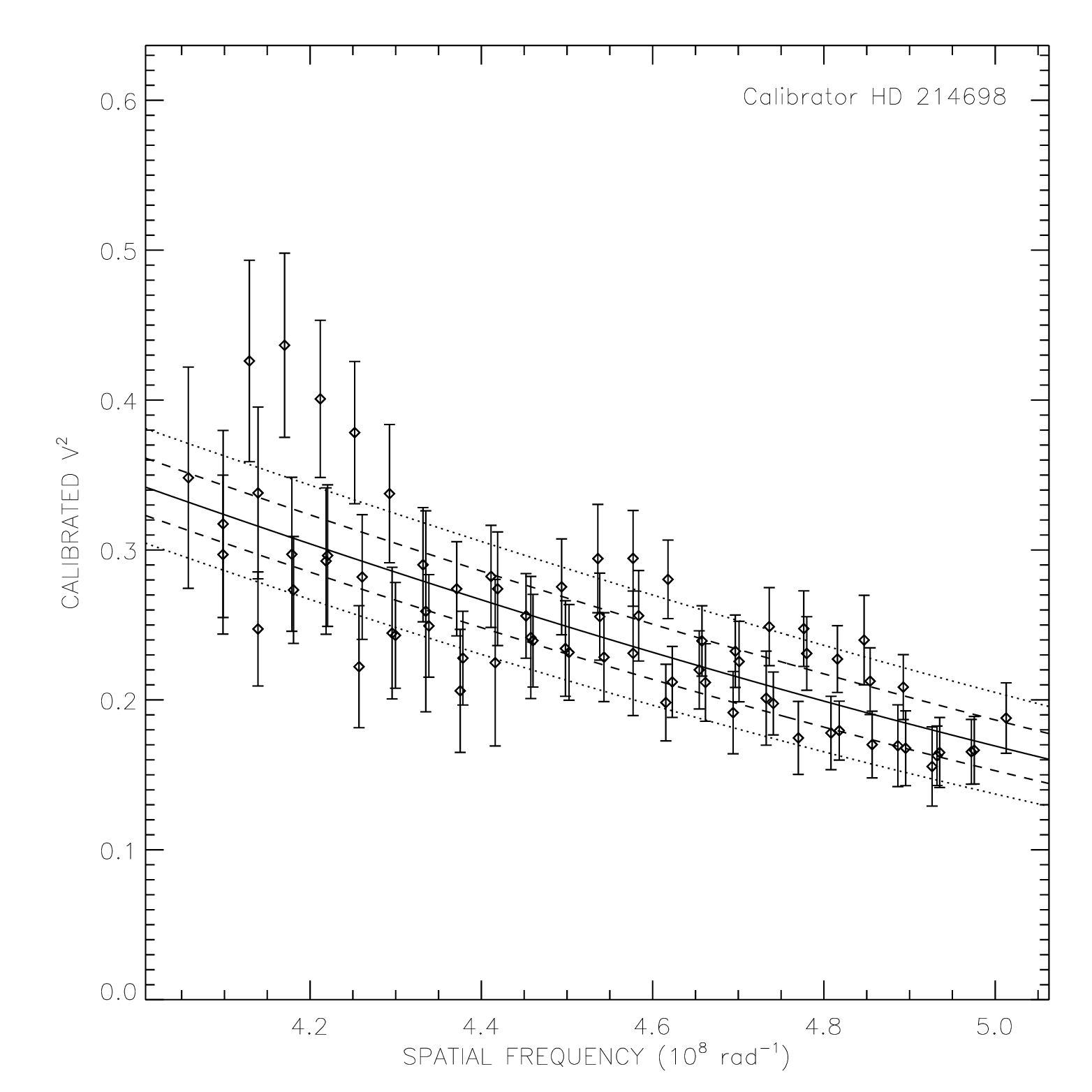}
\caption{$\theta_{\rm LD}$ fit for HR 8799 on 2011 Sep 30. The symbols are the same as in Figure \ref{HD218396_072510}.}
  \label{HD218396_093011}
\end{figure}

\begin{figure}[h]
\includegraphics[width=1.0\textwidth]{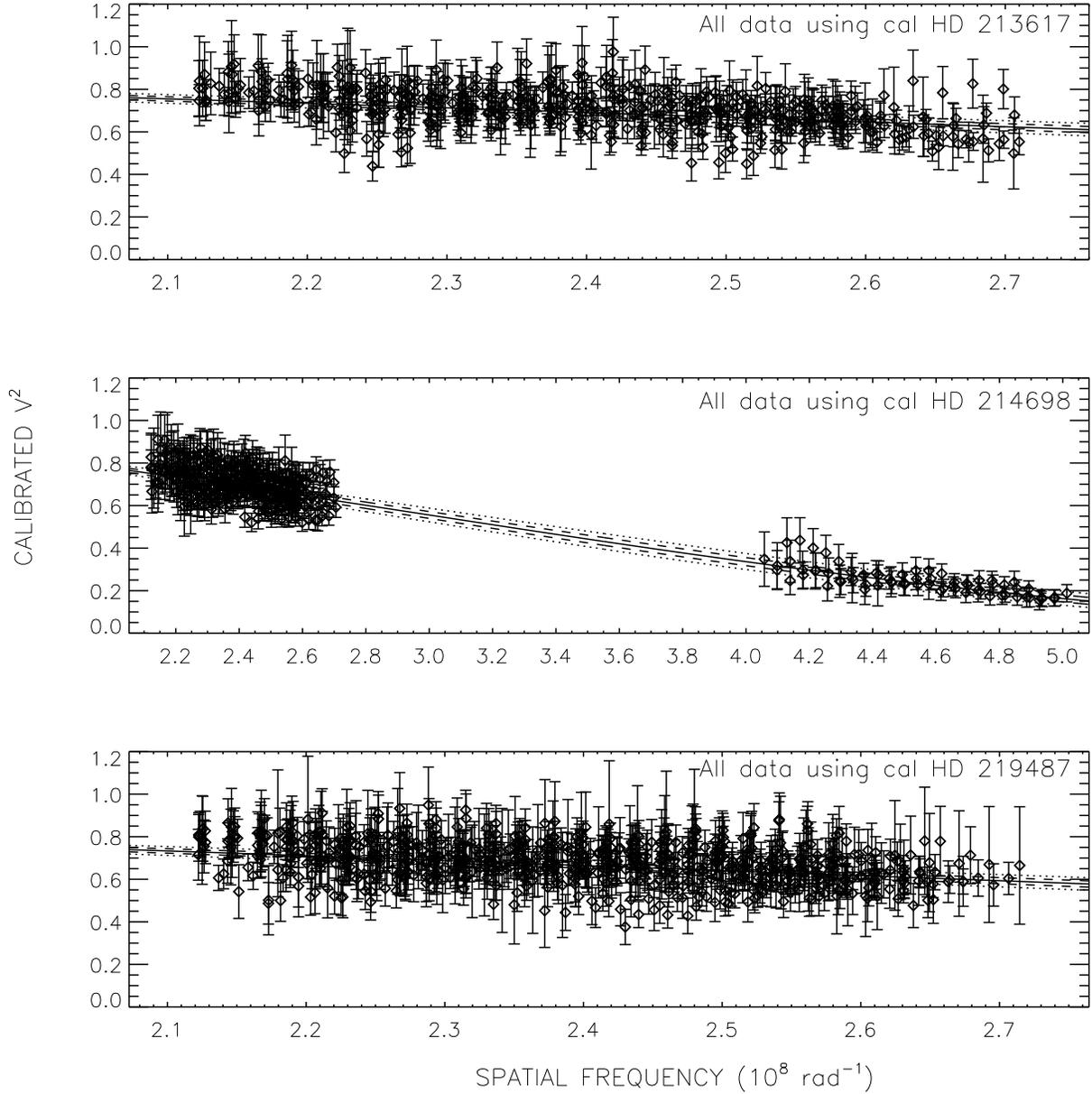}
\caption{$\theta_{\rm LD}$ fit for HR 8799 by calibrator. The symbols are the same as in Figure \ref{HD218396_072510}.}
  \label{HD218396_bycal}
\end{figure}

\begin{figure}[h]
\includegraphics[width=1.0\textwidth]{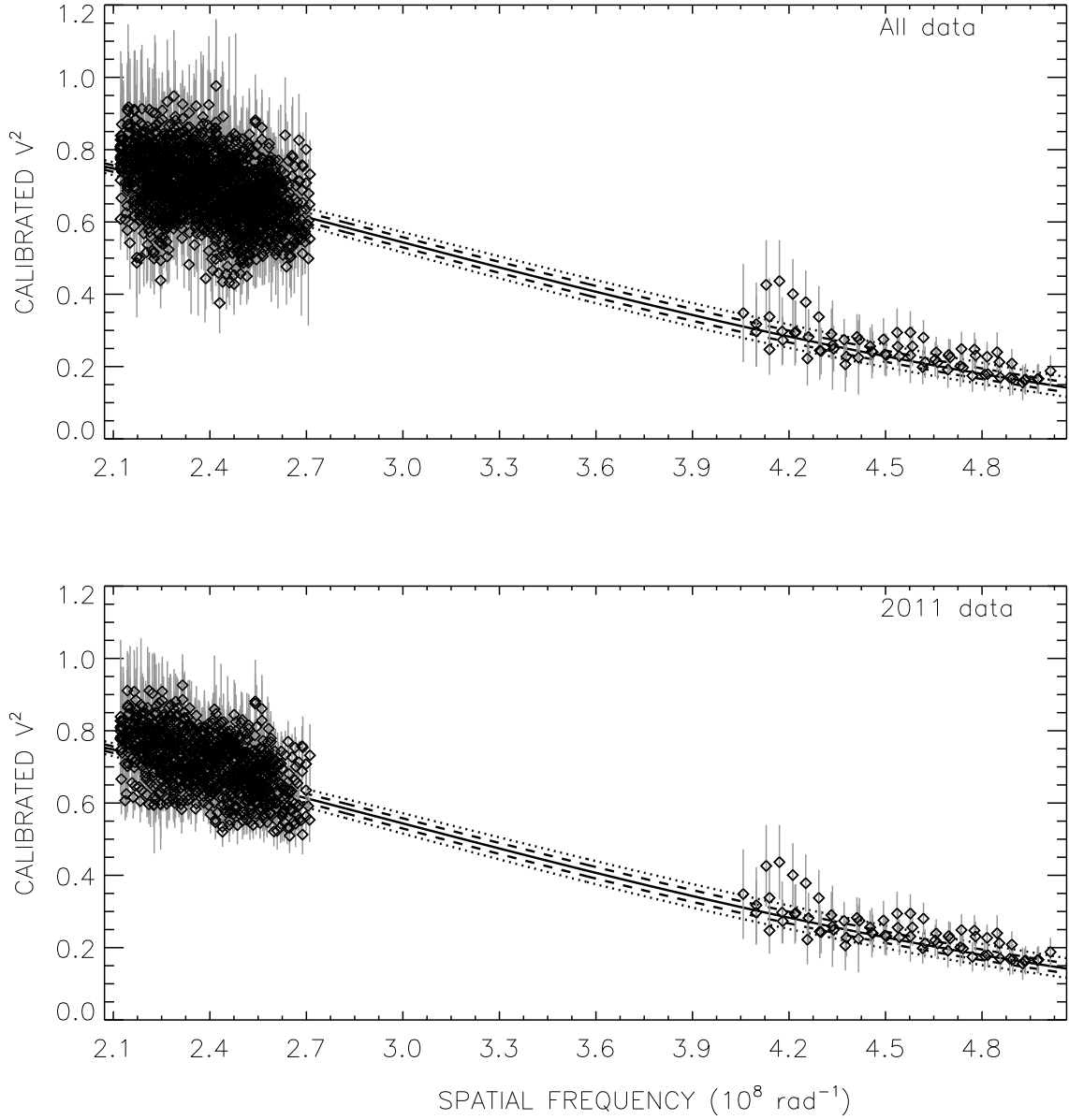}
\caption{$\theta_{\rm LD}$ fit for HR 8799 using all calibrators. The symbols are the same as in Figure \ref{HD218396_072510}.}
  \label{HD218396_all}
\end{figure}

The LD coefficient was obtained from \citet{2011AandA...529A..75C} after adopting the effective temperature ($T_{\rm eff}$) and surface gravity (log~$g$) values from \citet{1999AandA...352..555A}, which were 7586 K and 4.35, respectively. The resulting UD and LD angular diameters are listed in Table \ref{results}. Figures \ref{HD218396_072510} through \ref{HD218396_bycal} show the LD diameter fits for HR 8799 by night and by calibrator and Figure \ref{HD218396_all} shows all the data combined.
%The $T_{\rm eff}$ used to determine $\mu_{\lambda}$ has little effect on the final $\theta_{\rm LD}$: if $T_{\rm eff}$ varies by 500 K in either direction, $\mu_{\lambda}$ changed by 0.02 so we assigned $\mu_{\lambda}$ and error of 0.02. The final angular diameter is similarly little affected by the choice of $\mu_{\lambda}$: a 10$\%$ change in the $\mu_{\lambda}$ leads to a change in the measured $\theta_{\rm LD}$ of less than $0.6\%$, which is well within the 2$\%$ error bar of the final value. 

\begin{figure}[h]
\includegraphics[width=0.80\textwidth, angle=90]{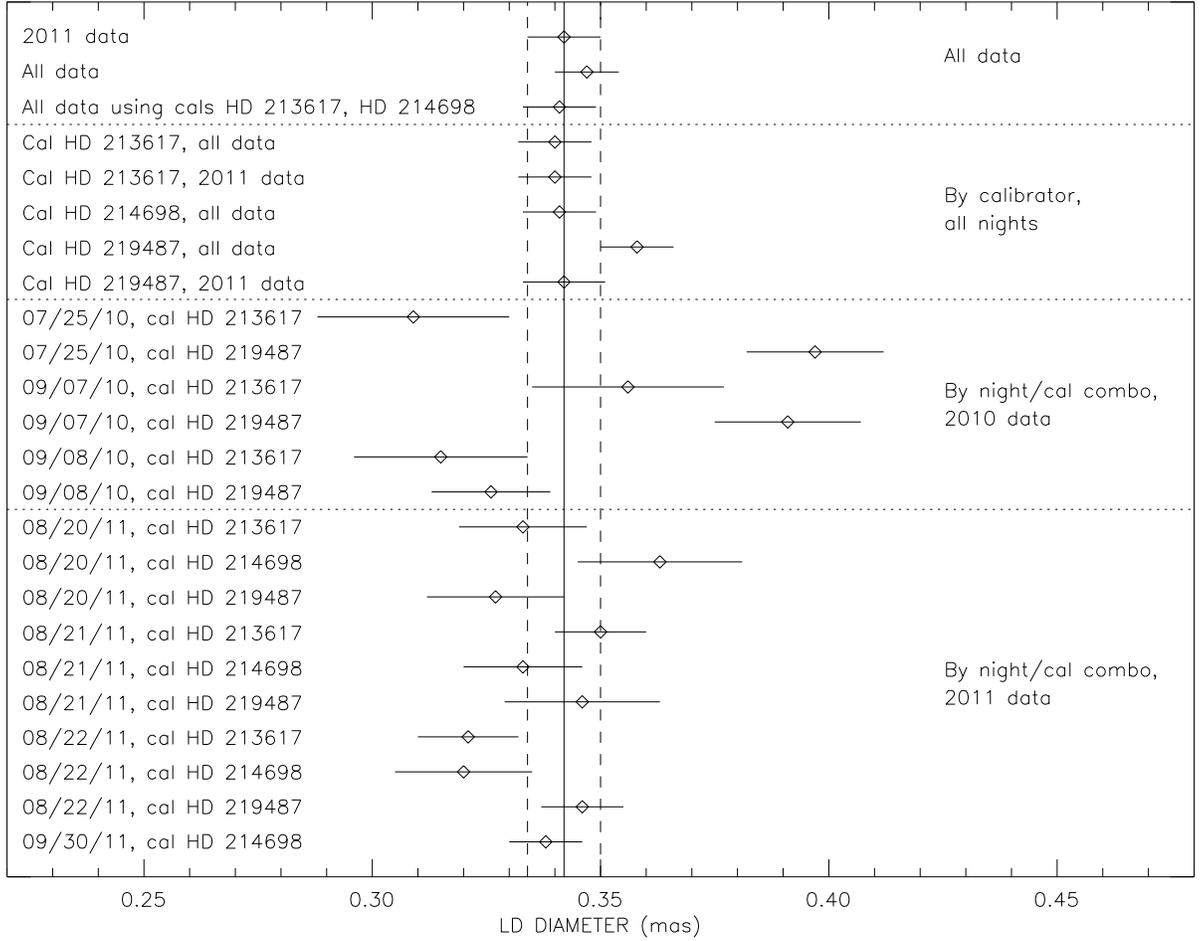}
\caption{A comparison of $\theta_{\rm LD}$ fits by individual night and calibrator. The vertical solid line represents the final adopted $\theta_{\rm LD}$ (0.342$\; \pm \;$0.008 mas, the top line) and the vertical dashed lines are the errors in that fit. Table \ref{hr8799_diams} lists the numerical values. Note the reduced scatter in the 2011 data versus the 2010 data.}
  \label{HD218396_diams}
\end{figure}

Table \ref{hr8799_diams} lists the resulting angular diameter fits for each night using each calibrator, and Figure \ref{HD218396_diams} shows a graphical version of the results. Though there is some scatter in the diameter fit from each night/calibrator combination, the scatter from the 2011 data is less pronounced than from 2010 data, which is seen in the diameter fits shown in Figures \ref{HD218396_082011} through \ref{HD218396_093011}. In particular, for observations obtained in 2010, the data exhibit sinusoidal-like variations about the best angular diameter fit; the variations are not seen in 2011 data. This is almost certainly due to coating asymmetries between CHARA Array telescopes (particularly overcoated silver versus aluminum) that were present in 2010 but removed in 2011. Attempts were made to search for polarization effects that would cause these residuals, but no conclusive evidence was found. Visibilities were, however, never measured in full Stokes parameters in a configuration that showed these residuals. 

Because the 2011 data show none of the sinusoidal residuals, our final $\theta_{\rm LD}$ incorporates the data from 2011 only. This yielded an angular diameter of 0.341$\pm$0.008 mas. When the data from all nights and using all calibrators are fit together, $\theta_{\rm LD}$ is 0.347$\pm$0.007 mas, which is within the uncertainty of the adopted value. 

The uncertainty for $\theta_{\rm LD}$ was calculated as described in the supplemental material from \citet{2011Sci...332..216D}: for each one of 10,000 Monte-Carlo simulated Gaussian distributions, we calibrated the instrumental visibilities and fit an angular diameter to the calibrated data using a least-squares minimization. We accounted for random errors by adding random numbers generated from the empirical covariance matrix scaled by the reduced $\chi^2$ of the original fit, and then repeated the procedure. The final uncertainty was the resulting standard deviation of the total distribution.

We observed HR 8799 with multiple calibrator stars every night except one to check on the behavior of the calibrators themselves. For example, when three calibrator stars were used, we used calibrator 1 to determine the angular diameters of calibrator 2 and calibrator 3 to make sure the stars were reliable. The calibrator HD 214698 shows the largest scatter, which is expected because it is the smallest calibrator of the three. There will naturally be more uncertainty when measuring its calibrated visibilities using the other two larger stars. The scatter seen in the night-to-night measurements is expected because the stars are small and very nearly unresolved.
%Figure \ref{cal_diams} presents the results of these fits.

\subsection{Stellar Radius, Luminosity and Effective Temperature}

At a distance of $39.4\pm1.1$ pc \citep{2007hnrr.book.....V}, HR 8799's $\theta_{\rm LD}$ of $0.342\pm0.008$ mas corresponds to a stellar radius\footnote{We define 1 solar radius to be $6.960\times10^{10}$ cm, consistent with an average of previous measurements \citep{2012ApJ...750..135E}.} of $1.44\pm0.06$ $R_\odot$, corresponding to a precision of 4$\%$. We note that this radius is 8$\%$ larger than the radius inferred from photometric and temperature considerations in \cite{1999AJ....118.2993G} and subsequently adopted in recent asteroseismic
analyses \citet[e.g.,][]{2010MNRAS.405L..81M,2010MNRAS.406..566M}; only 1$\%$ of this discrepancy can be attributed to the pre-revised Hipparcos distance used in the calculation of \cite{1999AJ....118.2993G}. 

In order to determine the luminosity ($L$) and $T_{\rm eff}$ of
HR 8799, we first constructed its photometric energy distribution (PED) from the averages (when multiple measurements were available) of Johnson $UBV$ magnitudes, Str\"omgren $uvby$ magnitudes, and 2MASS $JHK$ magnitudes, as published in \citet{1966AJ.....71..709C}, \citet{1968AJ.....73...84B},
\citet{1968QB4.G85n137....}, \citet{1986EgUBV........0M}, \citet{1986IBVS.2943....1S}, \citet{1990AandAS...84...29M}, \citet{1998AandAS..129..431H}, \citet{Gezari1999}, \citet{2000AandA...355L..27H}, \citet{2003tmc..book.....C}, and \citet{1983AandAS...54...55O, 1993AandAS..102...89O, 1994AandAS..106..257O}.
Table \ref{photometry} summarizes the adopted magnitudes; the assigned uncertainties for the 2MASS infrared measurements are as reported, and for optical measurements are standard deviations of multiple measurements. The assigned uncertainties of the optical measurements therefore account for the low amplitude optical variability ($\pm 0.02$ mag) observed for this variable star \citep{1999MNRAS.303..275Z}. 

\begin{figure}[h]
\includegraphics[width=1.0\textwidth]{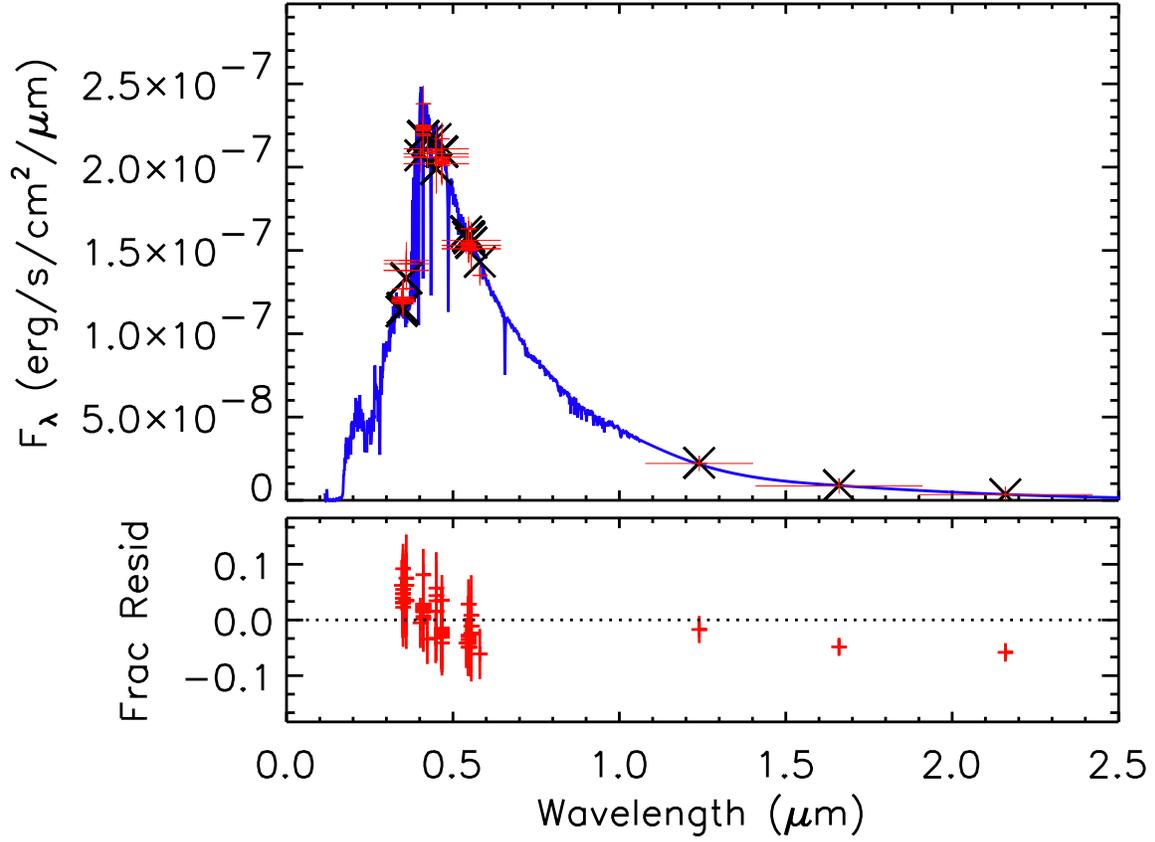}
\caption{HR 8799 PED fit. \emph{Upper panel:} The solid-line spectrum is a F0 V spectral template from \citet{1998PASP..110..863P}. The crosses indicate photometry values from the literature. The horizontal bars represent bandwidths of the filters used. The X-shaped symbols show the flux value of the spectral template integrated over the filter transmission. \emph{Lower panel:} The crosses are the residuals around the fit in fractional flux units of photometric uncertainty. For more details, see section 4.}
  \label{HD218396_sed}
\end{figure}

The bolometric flux ($F_{\rm BOL}$) of HR 8799 was determined by finding the best fit (via $\chi^2$ minimization) stellar spectral template from the flux-calibrated stellar spectral atlas of \citet{1998PASP..110..863P}. This best PED fit allows for extinction, using the wavelength-dependent reddening relations of \citet{1989ApJ...345..245C}. The best fit was found using a F0 V template with an assigned temperature of $7211 \pm 90$ K, an extinction of $A_{\rm V}$ = 0.00 $\pm$ 0.01 mag, and a $F_{\rm BOL}$ of $1.043\pm0.012 \times 10^{-7} \mathrm{\; erg \; s^{-1} \; cm^{-2}}$ (a 1.1$\%$ precision). Figure \ref{HD218396_sed} shows the best fit and the results are listed in Table \ref{results}. To check for possible systematic biases in our adopted prescription, we also fit the PED using synthetic Kurucz spectral models\footnote{Available to download at http://kurucz.cfa.harvard.edu.}, assuming no extinction. The resulting $F_{\rm BOL}$ estimate is consistent to within the 1.1$\%$ error reported above, further validating our technique and measured $F_{\rm BOL}$.

The bolometric flux was then combined with the distance to HR 8799 to estimate its luminosity ($L = 4 \pi$ d$^2 F_{\rm BOL}$), which yielded a value of $5.05 \pm 0.29$ $L_\odot$. The uncertainty in the luminosity (6$\%$) is predominantly set by the uncertainty in the distance. The $F_{\rm BOL}$ was also combined with the star's $\theta_{\rm LD}$
to determine its effective temperature by inverting the relation,
\begin{equation}
F_{\rm BOL} = {1 \over 4} \theta_{\rm LD}^2 \sigma T_{\rm eff}^4,
\end{equation}
where $\sigma$ is the Stefan-Bolzmann constant. This produces an effective temperature of $7203 \pm 87$ K, determined to a precision of 1$\%$. Because $\mu_{\rm LD}$ is chosen using a given $T_{\rm eff}$, we used our new $T_{\rm eff}$ value to select $\mu_{\rm LD}$ and iterated. $\mu_{\rm LD}$ increased by only 0.01 to 0.49$\pm$0.02, $\theta_{\rm LD}$ increased by 0.001 mas to 0.342$\pm$0.008 mas, and $T_{\rm eff}$ decreased by 10 K to 7193 K. The very slight change in $\theta_{\rm LD}$ did not affect the radius calculation at all. 

We note that this $T_{\rm eff}$ is nearly identical to the assigned temperature of the best fit stellar templates used to calculate $F_{\rm BOL}$. However, our interferometric $T_{\rm eff}$ is significantly less model dependent, and it is important to verify the accuracy of the $T_{\rm eff}$ predicted by the PED. Moreover, our measured $T_{\rm eff}$ is also consistent with the detailed spectral type classification by \citet{1999AJ....118.2993G} of kA5 hF0 mA5 v $\lambda$ Boo, following the spectral-type temperature scale for dwarf stars assembled in \citet{2007AJ....134.2340K}. The hydrogen lines are better tracer of the stellar $T_{\rm eff}$ (hF0; 7200 K); the metal lines suggest a higher $T_{\rm eff}$ (mA5; 8200 K) only because its atmosphere is slightly metal depleted.

A potential bias in the size measurement of any A-type star is oblateness caused by rapid rotation. \citet{2007AandA...463..671R} measured a $v \sin i$ of 49 km s$^{\rm -1}$ and assuming the planets orbit in the same plane as stellar rotation, the actual rotation velocity will increase from 49 to 104 km s$^{\rm -1}$ when the 28$^\circ$ inclination is taken into account \citep{2011ApJ...741...55S}. For an A5 V star with an approximate mass of 2.1 $M_\odot$ \citep{2000asqu.book.....C}, our measured radius of 1.44 $R_\odot$, and the relation between $M$, oblateness, and $v \sin i$ described in \citet{2006ApJ...637..494V}, the predicted oblateness of HR 8799 is only $\sim$2$\%$ and thus within the 1-$\sigma$ errors in $\theta_{\rm LD}$. Using the F0 V spectral type, which is a closer fit to our measured $T_{\rm eff}$, the estimated mass is 1.6 $M_\odot$ and the predicted oblateness is 2.6$\%$, still a small effect.

\section{Discussion}

\subsection{Habitable Real-Estate of HR 8799}
HR 8799 is currently the only directly imaged multiple planet system and there may be smaller, more Earth-like planets orbiting the star or even moons orbiting the imaged planets that have not yet been detected. This would present the possibility of life if the planets were in the habitable zone (HZ) of the star, so we used our new precise measurements to improve the estimate of the system's HZ. We used the following equations from \citet{2006ApJ...649.1010J}:
\begin{equation}
S_{b,i}(T_{\rm eff}) = (4.190 \times 10^{-8} \; T_{\rm eff}^2) - (2.139 \times 10^{-4} \; T_{\rm eff}) + 1.296
\end{equation}
and 
\begin{equation}
S_{b,o}(T_{\rm eff}) = (6.190 \times 10^{-9} \; T_{\rm eff}^2) - (1.319 \times 10^{-5} \; T_{\rm eff}) + 0.2341
\end{equation}
where $S_{b,i}$($T_{\rm eff}$) and $S_{b,o}$($T_{\rm eff}$) are the critical fluxes at the inner and outer boundaries in units of the solar constant. The inner and outer physical boundaries $r_{i,o}$ in AU were then calculated using
\begin{equation}
r_i = \sqrt{ \frac{L/L_\odot}{S_{b,i}(T_{\rm eff})} } \; \; \; \; \; {\rm and} \; \; \; \; \; r_o = \sqrt{ \frac{L/L_\odot}{S_{b,o}(T_{\rm eff})} }.
\end{equation}
These equations assume the HZ is dependent on distance only and do not take other effects, such as tidal heating, into account. The inner boundary is the limit where a runaway greenhouse effect would evaporate any surface water while the outer boundary is the limit where a cloudless atmosphere could maintain a surface temperature of 273 K. 
\citet{2006ApJ...649.1010J} note that these equations produce a conservatively small HZ and the actual HZ may be wider.

We obtained habitable zone boundaries of 1.62 AU and 3.32 AU. HR 8799's planets have semimajor axes of 14.5 to 68 AU \citep{2008Sci...322.1348M, 2010Natur.468.1080M}. There is no chance the planets orbit anywhere near the habitable zone unless they are highly eccentric, which is a configuration more likely to be unstable. 

\subsection{Effective Abundance of the $\lambda$ Boo Stars, and HR 8799}

The accurately determined stellar properties of HR 8799 also allow for detailed comparisons with stellar evolutionary models from which a mass and age can be inferred. However, correctly interpreting masses and especially ages depends critically upon knowing the internal abundances of this peculiar abundance star. As noted in the Introduction, the atmospheric abundances of Fe-peak elements are distinctly subsolar ([Fe/H] $\sim -0.4$ dex) while the abundances of C, N, O, and S are essentially solar. Because these surface abundances may not directly trace the internal abundances, it is unclear which, if any, of the available uniformly scaled abundance models to adopt for these comparisons; we subsequently refer to the abundance of the uniformly scaled model that predicts properties most consistent with observational constraints as the \textit{effective abundance}. 

As an example of the effect on the inferred ages, comparisons of HR 8799 with models that assume subsolar abundances throughout yield ages of $1.05 \pm 0.26$ Gyr for [Fe/H]= $-0.27$ or $1.71 \pm 0.18$ Gyr for [Fe/H] = $-0.43$, while those that assume near-solar abundances ([Fe/H] = $+0.05$) yield an age of $< 0.1$ Gyr; the specific physical assumptions used in these models are described in Section 4.3. Obviously, the polyabundic atmospheres of $\lambda$ Boo stars and their uncertain internal abundance compromises the validity of these comparisons and the inferred values. Nevertheless, we present the available observational evidence for constraining the metallicity of $\lambda$ Boo stars, and HR 8799 in particular, and suggest the effective abundance of HR 8799, and perhaps all $\lambda$ Boo stars, is near solar.

The observationally favored theory to explain the $\lambda$ Boo phenomenon, originally proposed by \citet{1990ApJ...363..234V}, is that the atmospheres have been polluted by the accretion of Fe-peak depleted gas. Depletion is believed to occur because of grain formation; a similar depletion of Fe-like elements has been observed in interstellar clouds \citep{1974ApJ...193L..35M}. The grains that themselves accrete Fe-elements are consequently inhibited from accreting on to the star because of the stronger radiation pressure they experience. It is not clear why this accretion onto the grains occurs, or where the accreting material comes from. It may be interstellar, but in many cases appears to be circumstellar. As noted by \citet{2002AJ....124..989G}, all four $\lambda$ Boo stars within 40 pc (including HR 8799) exhibit an infrared excess that is interpreted as the presence of circumstellar dust, while only $\sim 18 \%$ of non-$\lambda$ Boo A-type stars exhibit such excesses. The apparent depletion of these elements requires only relatively low accretion rates ($10^{-13}$ M$_\odot$ yr$^{-1}$), because of the shallow convective zones in A-type stars \citep{1991ApJ...372L..33C}. Accretion at these rates will quickly establish abundance anomalies within a few Myr, but these anomalies will likewise disappear in as little as 1 Myr once the accretion has terminated \citep{1993ApJ...413..376T}; the interesting implication is that all $\lambda$ Boo stars are either currently accreting, or have only recently terminated their accretion. If this favored theory is correct, it suggests that the depleted Fe-peak abundances do not represent the internal abundances of these stars. Given this and the typical limiting main-sequence lifetime of $\sim$2 Gyr for these intermediate mass Population I stars\footnote{Studies have shown that stars currently forming in open clusters are generally metal-rich \citep[e.g.,][]{2009AandA...493..309S}.}, one would expect their internal abundance to be on average close to solar.

We note that an alternative mechanism to explain the $\lambda$ Boo phenomenon proposed by \citet{1986ApJ...311..326M} suggests that the depletion of Fe-peak elements is a result of diffusion and mass loss. In this case, $\lambda$ Boo stars are much closer to the end of their main-sequence lifetimes.  To be effective, the star would need to be losing mass at a rate of order $10^{-13}$ M$_\odot$ yr$^{-1}$ for a few times $10^8$ yr and thus implies that the $\lambda$ Boo phenomenon be restricted to the end of the main-sequence evolutionary phase for A stars. This predicted timescale is difficult to reconcile with the discovery of very young $\lambda$ Boo stars in the Orion OB1 associations (age $< 10$ Myr), and the lack of any $\lambda$ Boo stars in intermediate age open clusters \citep{2002AJ....124..989G}. Evidence such as the higher fraction of these stars with circumstellar debris disks, noted above, instead suggest that the $\lambda$ Boo phenomenon is more common among young A stars. If true, this would strengthen the case that the internal abundances of these Population I stars, at least on average, should be close to solar. In particular, we note that the existence of $\lambda$ Boo stars in the Orion star forming region implies that their primordial abundances are solar, consistent with other stars in this cluster \citep{2012AandA...539A.143N}. If the depletion of Fe-like elements is restricted to only the outer atmospheres, then one can conclude that in these cases the internal abundances should likewise be solar. 

\begin{figure}[h]
\includegraphics[width=1.0\textwidth]{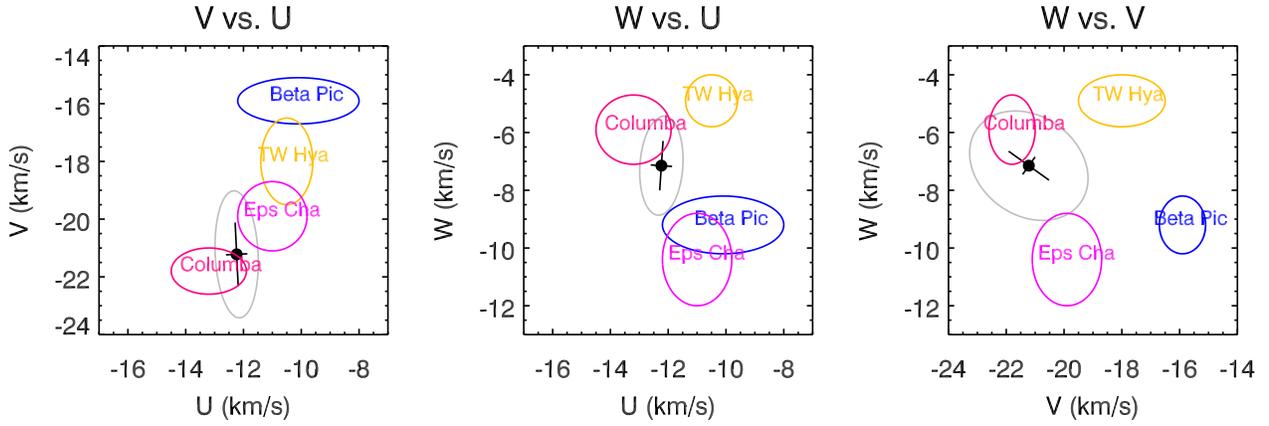}
\caption{The $UVW$ space velocities of HR 8799 (black point) with 1-$\sigma$ (error bars) and 2-$\sigma$ (gray ellipse) errors.  Also plotted are the $UVW$ space velocities and 1-$\sigma$ errors (as ellipses) of four young stellar associations with similar kinematics, as taken from \citet{2008hsf2.book..757T}. HR 8799 matches the space velocity of the roughly 30 Myr old Columba association to 1.2$\sigma$.}
 \label{spacemotions} 
\end{figure}

As described in \citet{2008Sci...322.1348M}, the space motions of HR 8799 are consistent with those of young stars in the solar neighborhood. We investigated this further by comparing the $UVW$ space motions of HR 8799 to those of nearby moving groups assembled in \citet{2008hsf2.book..757T}; kinematic association with a moving group would not only help constrain the age of the system but also the primordial abundance of the star. We calculated {\it UVW} space motions using HR 8799's \emph{HIPPARCOS} distance and parallax \citep{2007hnrr.book.....V}, and a radial velocity of $-$12.6$\pm$1.3 km s$^{-1}$ from \citet{2000AandAS..142..217B}. This yielded velocities of {\it U =} $-$12.24$\pm$0.37 km s$^{-1}$, {\it V =} $-$21.22$\pm$1.10 km s$^{-1}$, and {\it W =} $-$7.15$\pm$0.86 km s$^{-1}$; these are all within $\sim$1 km s$^{-1}$ of the values reported in \citet{2008Sci...322.1348M}. Figure \ref{spacemotions} illustrates the space motion of HR 8799 relative to that of several kinematically similar moving groups after adopting moving group velocities and uncertainties from \citet{2008hsf2.book..757T}. Although the space motion of HR 8799 is not consistent to within 1$\sigma$ of any of these groups, it is consistent to within 3$\sigma$ of two of these groups: Columba (1.2$\sigma$), and $\epsilon$ Cha (2.2$\sigma$). We thus confirm the report by \citet{2008Sci...322.1348M} of the potentially youthful kinematics of HR 8799, which favors but can not confirm an age less than $\lesssim 1$ Gyr, and likely a solar abundance consistent with many of these stars.

As described in the Introduction, there are two other aspects of HR 8799 that argue for a young age for the star and system. The strong infrared excess of HR 8799 statistically favors an age less than $\sim 500$ Myr \citep{2009ApJ...697.1578G}, despite the challenge raised by \citet{2010MNRAS.405L..81M}. An even stronger youthful age restriction is imposed by the dynamical considerations of \citet{2010ApJ...721L.199M}. Although their conclusions require use of the predictions of poorly constrained planetary evolutionary models, even considerable uncertainties in the mass estimates require the age of the system to be less than a few 100 Myr, otherwise the planets would be too massive to be orbitally stable. The ages much less than $1$ Gyr for HR 8799 can only be reconciled if the adopted effective abundances are near solar.

Recently, \citet{2010MNRAS.406..566M} used the $\gamma$ Dor-type pulsations of HR 8799 to asteroseismically constrain the star's internal metallicity. They find a best fit internal metallicity of $-0.32 \pm 0.1$, assuming that the rotation axis of the star, which affects the rotation speed of the star, is somewhat highly inclined relative to our line of sight (50$^\circ$). This high inclination is inconsistent with the more face-on orientation of HR 8799's disk \citep{2009ApJ...705..314S} and the apparent orbital plane of its companions \citep{2009ApJ...694L.148L}, although these need not be coplanar. The asteroseismology analysis is compromised by the use of an assumed radius that is too small by 8$\%$ and a temperature that is too hot by 3$\%$ relative to the precisely determined values in this study. The measured $R$ and $T_{\rm eff}$ presented here will be valuable to constrain the model parameter space in future asteroseismic modeling efforts, which will possibly constrain the internal metallicity.

\citet{2010MNRAS.406..566M} do find possible -- though less likely -- solutions with internal abundances close to solar ([Fe/H] $= -0.12$) for closer to pole-on orientations; this is the most metal-rich metallicity reported in their study. Overall we find the results of this metallicity analysis to be inconclusive primarily because of the large uncertainty in the inclination of the star's rotation axis, which Moya et al. acknowledge is the limiting factor in their analysis. If anything, the result that they find some acceptable solutions with near solar metallicities and young ages corroborated previous asteroseimology studies of $\lambda$ Boo stars. Using densities inferred from stellar pulsations, \citet{1998AandA...335..533P} found the location of most $\lambda$ Boo stars showing $\delta$ Scuti pulsations on a plot of average density versus period to be consistent with the positions of $\delta$ Scuti stars with solar abundances. They interpret this as evidence that the low Fe-peak abundances are restricted to the surface of $\lambda$ Boo stars and do not represent the state of the interior composition. Although the variations in HR 8799 are compatible with $\gamma$ Dor pulsations, the large number of discovered hybrid $\delta$ Scu - $\gamma$ Dor pulsators \citep[e.g.,][]{2006CoAst.148...34R, 2010AN....331..989G} suggests that this argument could be applicable for all $\lambda$ Boo stars.

In summary, based on the observational evidence supporting the accretion of clean gas theory explaining the abundance pattern of $\lambda$ Boo stars, the existence of at least some $\lambda$ Boo stars in an OB Association, the evidence of youth for many $\lambda$ Boo stars (and especially HR 8799), the restriction of ages less than $\sim$2 Gyr for Population I stars, and the available asteroseismic constraints on internal metallicity, we conclude that the most appropriate \textit{effective} abundances for HR 8799, and quite possibly all $\lambda$ Boo stars, is near solar. Until more sophisticated evolutionary models that account for the polyabundic atmospheres and possibly interiors of these stars are developed, we are hopeful that $\lambda$ Boo stars in clusters or with lower mass companions will be discovered, which will enable improvements in both assigning effective abundances and tests of the validity of using uniformly scaled abundances. 

\subsection{Mass and Age of HR 8799 and Implications for Its Companions}

\begin{figure}[h]
\includegraphics[width=0.8\textwidth, angle=90]{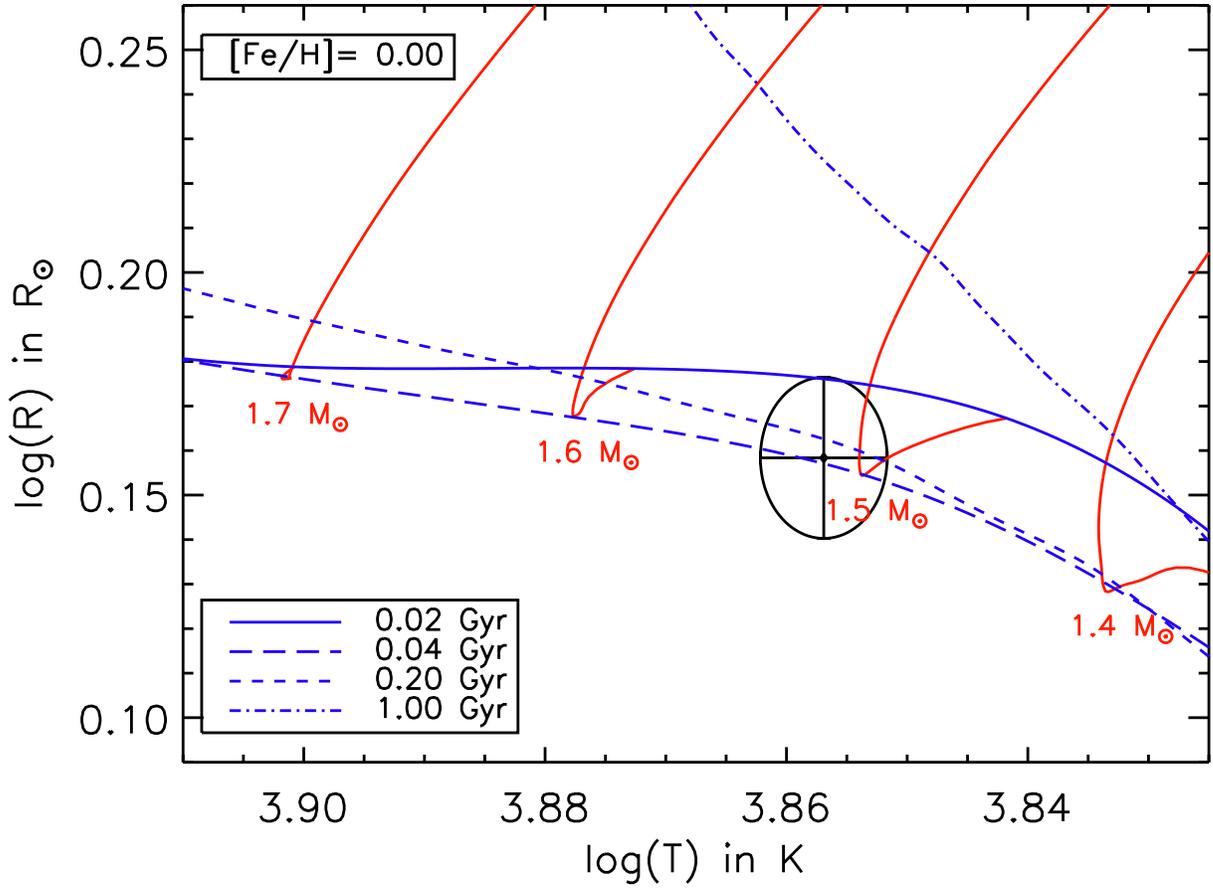}
\caption{Radius and temperature of HR 8799 plotted along with isochrones (blue lines) and mass tracks (solid red lines) from \citet{2004ApJS..155..667D}, with solar abundances. Note that the mass tracks predict that these stars are still gravitationally settling at 0.02 Gyr, reach their smallest size at 0.04 Gyr, and expand in size thereafter.}
 \label{HR_solar_ell} 
\end{figure}

Following the above arguments, we estimated the mass and age of HR 8799 using the stellar evolutionary models of the Yonsei-Yale group \citep[Y$^{\rm 2}$,][]{2001ApJS..136..417Y}, updated by \citet{2004ApJS..155..667D} to account for convective core overshoot. The models handle convection using mixing length prescription, adopting a mixing length of 1.7432 times the pressure scale height that is set by comparisons with a solar model. The model uses the solar abundances of \citet{1996ASPC...99..117G} and OPAL radiative opacities \citep{1995HiA....10..573R,1996ApJ...464..943I} for the interior and the \citet{1994ApJ...437..879A} opacities for the cooler, outer regions of the star. Additional information regarding the input physics assumed in these models can be found in \citet{2001ApJS..136..417Y} and \citet{2004ApJS..155..667D}.  However, since a set solar metallicity models is not provided, we followed the recommendation of the modelers themselves and generated a set of solar metallicity models by linearly interpolating isochrones and mass tracks between nearest the metallicity models of [Fe/H]=-0.27 dex and [Fe/H] = +0.05 dex (models x74z01 and x71z04 models, respectively). The solar metallicity models are illustrated in Figure \ref{HR_solar_ell}.

At HR 8799's temperature, stars are predicted to have fully contracted to their smallest size, or the zero age main sequence (ZAMS) at 40 Myr, and then begin expanding due to stellar evolutionary effects at a much slower rate afterwards. As such, HR 8799 is either contracting onto the ZAMS or expanding from it, and we use the 40 Myr isochrone as the upper limit for the age in the pre-main-sequence scenario and as the lower limit for the age in the post-main-sequence scenario. More specifically, we find that if HR 8799 is contracting onto the ZAMS, it has an age of $33^{+7}_{-13.2}$ Myr and a mass of $1.516^{+0.038}_{-0.024}$ $M_{\odot}$. If HR 8799 is expanding from the ZAMS, we find it to have an age of $90^{+381}_{-50}$ Myr and a mass of $1.513^{+0.023}_{-0.024}$ $M_{\odot}$. These masses and ages are found by interpolating between the solar metallicity models described above. The masses are consistent with the mass used in \citet{2008Sci...322.1348M}, which was 1.5$\pm$0.3 $M_\odot$. We remind the reader that these quoted errors on the star's mass and age are statistical and therefore do not take into account uncertainties in the metallicity and/or models themselves.

With an age of $\lesssim$ 0.1 Gyr, the companions that HR 8799 harbors are more likely to have planetary masses. As explained in \citet{2008Sci...322.1348M}, planetary mass objects could only be as bright as these observed companions are if they are young. That inferred age, and thus companion masses, depends critically upon adopted evolutionary model abundance. While near solar seems likely, even slightly subsolar abundances can give uncertainties in age that extend to above $\sim 500$ Myr, increasing the inferred companion masses by at least a factor of two. 

%The Y$^2$ models with an abundance closest to solar have [Fe/H] $=+0.05$, so are in fact slightly metal rich; the next closest models are computed with [Fe/H] $=-0.27$ and then $-0.43$. Figures \ref{HR_metalpoor} and \ref{HR_solar} illustrate the measured stellar radius and effective temperature of HR 8799 along with isochrones of the near solar models.  The datapoint lies slightly below the range of stellar radii predicted by the $+0.05$ model, although the observational error bars extend into the range of predicted sizes. 

%At HR 8799's temperature, stars are predicted to have fully contracted to their smallest size (i.e., zero age main sequence) at 40 Myr, and then begin expanding due to stellar evolutionary effects at a much slower rate afterwards. Adopting the smallest radius isochrone as the most likely age, and using the uncertainty in radius to assign an uncertainty in age, we found an age of $40^{+340}_{-20}$ Myr. The mass of HR 8799 was determined using the 40 Myr isochrone along with its measured temperature and uncertainty. This yielded a mass of 1.54$\pm$0.02 M$_\odot$. This is consistent with the mass used in \citet{2008Sci...322.1348M}, which was 1.5$\pm$0.3 $M_\odot$. It should be noted the error on the star's mass is statistical and given the uncertainty in metallicity and the models themselves, an error of 0.1 $M_\odot$ may be more realistic.

\section{Summary}
We measured the angular diameter of HR 8799 using the CHARA Array interferometer and used our new value of 0.342$\pm$0.008 mas to calculate the star's physical radius (1.44$\pm$0.06 $R_\odot$), luminosity (5.05$\pm$0.29 $L_\odot$), and effective temperature (7193$\pm$10 K) by combining our measurement with information from the literature. We used our $T_{\rm eff}$ measurement to determine the size of the habitable zone, which is well inside the orbits of any of the companions detected to date.

Based on a variety of techniques, we concluded that the most appropriate abundances for HR 8799 are close to solar. We combined our $R$ and $T_{\rm eff}$ values with $Y^{\rm 2}$ isochrones to estimate the star's mass and age in two scenarios: $1.516^{+0.038}_{-0.024}$ $M_{\odot}$ and $33^{+7}_{-13.2}$ if the star is contracting onto the ZAMS or $1.513^{+0.023}_{-0.024}$ $M_{\odot}$ and $90^{+381}_{-50}$ if it is expanding from it. In either case, this young age implies the imaged companions are planets and not brown dwarfs. The only case in which the companions would be close to brown dwarf mass is if the highest age of 471 Myr was the true case, and even then the masses would be on the exoplanet/brown dwarf cusp.

\acknowledgments

We would like to thank Gerard van Belle for his insight on the nature of HR 8799's pirouette through space. The CHARA Array is funded by the National Science Foundation through NSF grant AST-0606958 and by Georgia State University through the College of Arts and Sciences, and STR acknowledges partial support by NASA grant NNH09AK731. This research has made use of the SIMBAD database, operated at CDS, Strasbourg, France. This publication makes use of data products from the Two Micron All Sky Survey, which is a joint project of the University of Massachusetts and the Infrared Processing and Analysis Center/California Institute of Technology, funded by the National Aeronautics and Space Administration and the National Science Foundation.

%%%%%%%%%%%%%%%%%%%%%%% System Parameters %%%%%%%%%%%%%%%%%%%%%%%%%%%%
\begin{deluxetable}{ccll}
%\rotate
\tablewidth{0pc}
\tabletypesize{\scriptsize}
\tablecaption{HR 8799 System Parameters from the Literature. \label{lit_params}}

\tablehead{\colhead{$M_{\rm star}$} & \colhead{Age}   & \colhead{ }           & \colhead{ }  \\ 
           \colhead{($M_\odot$)}    & \colhead{(Myr)} & \colhead{Method Used} & \colhead{Reference} }
\startdata
1.47$\pm$0.30 & -- & spectral synthesis \& spectrophotometry & \citet{1999AJ....118.2993G} \\
-- & 50-1128 & H-R diagram placement & \citet{2001ApJ...546..352S} \\
%-- & $\lesssim$50 & IR excess of debris disk & \citet{2006ApJ...653..675S} \\
-- & 30 & stellar kinematics, H-R diagram & \citet{2004ApJ...603..738Z} \\
   &    & placement vs. isochrones &   \\
-- & 20-150 & Local Association membership & \citet{2006ApJ...644..525M} \\
-- & 590 & IR excess & \citet{2006ApJS..166..351C} \\
-- & 30 & IR excess, H-R diagram placement & \citet{2007ApJ...660.1556R} \\
1.5 & 30-160 & multiple methods & \citet{2008Sci...322.1348M} \\
1.2-1.6 & $\lesssim$100 Myr & dynamical stability analysis & \citet{2009MNRAS.397L..16G} \\
-- & $\lesssim$50 & observational data analysis & \citet{2009AandA...503..247R} \\
1.5 & $\lesssim$100 Myr & dynamical stability analysis & \citet{2010ApJ...710.1408F} \\
--  & $\sim$100 Myr & disk inclination & \citet{2010ApJ...721L.199M} \\
1.32-1.33$^{\rm a}$ & 1123-1623 & asteroseismology & \citet{2010MNRAS.405L..81M} \\
1.44-1.45$^{\rm b}$ & 26-430 & asteroseismology & \citet{2010MNRAS.405L..81M} \\
1.32 & 1126-1486 & asteroseismology & \citet{2010MNRAS.405L..81M} \\
-- & 30 & Columba Association membership & \citet{2010lyot.confE..42D} \\
-- & 30 & Columba Association membership & \citet{2011ApJ...732...61Z} \\
-- & 30-100 & direct imaging & \citet{2011ApJ...729..128C} \\
-- & 30-300 & atmospheric/evolution model fitting & \citet{2012ApJ...754..135M} \\
-- & 30-155 & dynamical stability analysis & \citet{2012ApJ...755...38S} \\
\enddata
\tablecomments{$^{\rm a}$Solution 1; $^{\rm b}$Solution 2}
\end{deluxetable}

\clearpage

%%%%%%%%%%%%%%%%%%%%%%% Calibrators %%%%%%%%%%%%%%%%%%%%%%%%%%%%%%%
\begin{deluxetable}{cccccccc}
\tablewidth{0pc}
%\rotate
%\tabletypesize{\scriptsize}
\tablecaption{Calibrator Star Properties.\label{cals}}
\tablehead{ \colhead{}   & \colhead{Spectral} & \colhead{$V$} & \colhead{$K$} & \colhead{E($B-V$)} & \colhead{$\theta_{\rm photometric}$} \\
            \colhead{HD} & \colhead{Type}     & \colhead{(mag)} & \colhead{(mag)} & \colhead{ }      & \colhead{(mas)} }
\startdata
213617 & F1 V & 6.42 & 5.58$\pm$0.02 & 0.03 & 0.288$\pm$0.006 \\
214698 & A2 V & 6.33 & 6.21$\pm$0.02 & 0.05 & 0.189$\pm$0.005  \\
219487 & F5 V & 6.60 & 5.54$\pm$0.02 & 0.05 & 0.304$\pm$0.006   \\
\enddata
\tablecomments{Spectral types are from SIMBAD; $V$ magnitudes are from \citet{Mermilliod} except for HD 214698, which is from \citet{1997ESASP1200.....P}. No errors were listed so we assigned errors of $\pm$0.01 mag; $K$ magnitudes are from \citet{2003tmc..book.....C}.}
\end{deluxetable}
  
\clearpage

%%%%%%%%%%%%%%%%%%% Calibrated Visibilities %%%%%%%%%%%%%%%%%%%%%

\begin{deluxetable}{ccccc}
\tablewidth{0pc}
\tablecaption{Calibrated Visibilities.\label{calib_visy}}

\tablehead{\colhead{ }    & \colhead{Calibrator} & \colhead{Spatial Freq}       & \colhead{ }     & \colhead{ }          \\
           \colhead{Date} & \colhead{Used}       & \colhead{(10$^8$ rad$^{-1}$)} & \colhead{$V^2$} & \colhead{$\sigma V^2$} }
\startdata
2010 Aug 25 & HD 213617 & 2.223 & 0.861 & 0.081	 \\
            &           & 2.245 & 0.768 & 0.069	 \\
            &           & 2.267 & 0.785 & 0.069	 \\
            &           & 2.289 & 0.807 & 0.076	 \\
            &           & 2.311 & 0.804 & 0.089	 \\
            &           & 2.332 & 0.729 & 0.094	 \\
            &           & 2.353 & 0.811 & 0.138	 \\
            &           & 2.375 & 0.844 & 0.146	 \\
            &           & 2.397 & 0.924 & 0.142	 \\
            &           & 2.419 & 0.976 & 0.135	 \\
            &           & 2.442 & 0.892 & 0.092	 \\
            &           & 2.464 & 0.834 & 0.066	 \\
            &           & 2.486 & 0.788 & 0.059	 \\
            &           & 2.507 & 0.764 & 0.057	 \\
            &           & 2.528 & 0.804 & 0.063	 \\
            &           & 2.550 & 0.742 & 0.059	 \\
            &           & 2.571 & 0.728 & 0.067	 \\
            &           & 2.593 & 0.713 & 0.084	 \\
            &           & 2.613 & 0.771 & 0.104	 \\
            &           & 2.634 & 0.840 & 0.120	 \\
            &           & 2.655 & 0.784 & 0.103	 \\
            &           & 2.677 & 0.827 & 0.096	 \\
            &           & 2.699 & 0.801 & 0.077	 \\

\enddata
\tablecomments{All data were taken using the S2-W2 baseline (177 m). Only a portion of this table is shown here to demonstrate its form and content. A machine-readable version of the full table is available on the online version of \emph{The Astrophysical Journal}.}
\end{deluxetable}

\clearpage

%%%%%%%%%%%%%%%%%%% HR 8799 Diams %%%%%%%%%%%%%%%%%%%%%
\begin{deluxetable}{lccc}
\tablewidth{0pc}
%\rotate
%\tabletypesize{\scriptsize}
\tablecaption{HR 8799 Angular Diameter Measurements.\label{hr8799_diams}}
\tablehead{\colhead{}     & \colhead{Calibrator} & \colhead{$\theta_{\rm UD}$} & \colhead{$\theta_{\rm LD}$} \\
           \colhead{Date} & \colhead{HD}         & \colhead{(mas)}             & \colhead{(mas)}  }
\startdata
2010 Aug 25 & 213617 & 0.297$\pm$0.021 & 0.309$\pm$0.021 \\
            & 219487 & 0.381$\pm$0.015 & 0.397$\pm$0.016 \\
            & All    & 0.357$\pm$0.016 & 0.372$\pm$0.016  \\
2010 Sep 07 & 213617 & 0.342$\pm$0.015 & 0.356$\pm$0.014 \\
            & 219487 & 0.375$\pm$0.010 & 0.391$\pm$0.010 \\
            & All    & 0.359$\pm$0.009 & 0.373$\pm$0.010  \\
2010 Sep 08 & 213617 & 0.302$\pm$0.018 & 0.315$\pm$0.019 \\
            & 219487 & 0.313$\pm$0.012 & 0.326$\pm$0.013 \\
            & All    & 0.308$\pm$0.011 & 0.321$\pm$0.012  \\
2011 Oct 20 & 213617 & 0.320$\pm$0.013 & 0.333$\pm$0.014 \\
            & 214698 & 0.349$\pm$0.017 & 0.363$\pm$0.018 \\
            & 219487 & 0.314$\pm$0.015 & 0.327$\pm$0.015 \\
            & All    & 0.331$\pm$0.009 & 0.345$\pm$0.010  \\
2011 Oct 21 & 213617 & 0.336$\pm$0.010 & 0.350$\pm$0.010 \\
            & 214698 & 0.320$\pm$0.012 & 0.333$\pm$0.013 \\
            & 219487 & 0.332$\pm$0.016 & 0.346$\pm$0.017 \\
            & All    & 0.330$\pm$0.019 & 0.343$\pm$0.021  \\
2011 Oct 22 & 213617 & 0.308$\pm$0.010 & 0.321$\pm$0.011 \\
            & 214698 & 0.308$\pm$0.014 & 0.320$\pm$0.015 \\
            & 219487 & 0.332$\pm$0.008 & 0.346$\pm$0.009 \\
            & All    & 0.328$\pm$0.023 & 0.341$\pm$0.022  \\
2011 Sep 30 & 214698 & 0.323$\pm$0.009 & 0.338$\pm$0.008 \\
\cline{1-4}
All dates          & 213617 & 0.327$\pm$0.008 & 0.341$\pm$0.008  \\
Aug 2011 data only &        & 0.326$\pm$0.008 & 0.340$\pm$0.008  \\
All dates          & 214698 & 0.327$\pm$0.009 & 0.341$\pm$0.010  \\
All dates          & 219487 & 0.343$\pm$0.008 & 0.358$\pm$0.008  \\
Aug 2011 data only &        & 0.328$\pm$0.009 & 0.342$\pm$0.009  \\
\cline{1-4}
%SED fit            &        &                 & 0.341$\pm$0.009  \\
All dates          & All    & 0.333$\pm$0.007 & 0.347$\pm$0.007  \\
Final fit: 2011 data & All    & 0.327$\pm$0.007 & 0.342$\pm$0.008  \\
%\cline{1-4}
%All dates          & HD 213617 \& &  &  \\
%                   & HD 214698 only & 0.327$\pm$0.008 & 0.341$\pm$0.008  \\
\enddata
\tablecomments{Figure \ref{HD218396_diams} shows a graphical view of these values. Data were obtained using the S2-W2 baseline at 177 m for all nights except for 2011 Sep 30, which were obtained using the S1-E1 baseline at 331 m.}
\end{deluxetable}

\clearpage

%%%%%%%%%%%%%%%%%%%%%%% Photometry %%%%%%%%%%%%%%%%%%%%%%%%%%%%%%%

\begin{deluxetable}{ccc}
\tablewidth{0pc}
\tablecaption{HR 8799 Photometry\label{photometry}}
\tablehead{ \colhead{}   & \colhead{Average} & \colhead{No. of }  \\
           \colhead{Band} & \colhead{(mag)} & \colhead{measurements} }
\startdata
$U$    & 6.191$\pm$0.016    & 2 \\
$B$    & 6.235$\pm$0.016    & 2 \\
$V$    & 5.980$\pm$0.016    & 2 \\
$u$   & 7.471$\pm$0.025    & 7 \\
$v$   & 6.468$\pm$0.025    & 7 \\
$b$   & 6.143$\pm$0.027    & 7 \\
$y$   & 5.962$\pm$0.026    & 7 \\
$J$    & 5.383$\pm$0.027    & 1 \\
$H$    & 5.280$\pm$0.018    & 1 \\
$K$    & 5.240$\pm$0.018    & 1 \\
\enddata
\end{deluxetable}

\clearpage

%%%%%%%%%%%%%%%%%%%%%%% Results %%%%%%%%%%%%%%%%%%%%%%%%%%%%%%%
\begin{deluxetable}{lcl}
\tablewidth{0pc}
%\tabletypesize{\scriptsize}
\tablecaption{HR 8799 Stellar Parameters.\label{results}}
\tablehead{ \colhead{Parameter} & \colhead{Value} & \colhead{Reference} }
\startdata

$V$ magnitude     & 5.98$\pm$0.01 & \citet{Mermilliod} \\
$K$ magnitude     & 5.24$\; \pm \;$0.02 & \citet{2003tmc..book.....C} \\
$\pi$ (mas) & 25.38$\; \pm \;$0.70 & \citet{2007hnrr.book.....V} \\
Distance (pc) & 39.40$\; \pm \;$1.09 & \citet{2007hnrr.book.....V} \\
$\mu_{\lambda}$ & 0.49$\pm$0.02 & \citet{2011AandA...529A..75C} \\
\cline{1-3}
$A_{\rm V}$       & 0.00$\; \pm \;$0.01 & This work; PED fit \\
$F_{\rm BOL}$ (10$^{-7}$ erg s$^{-1}$ cm$^{-2}$) & 1.043$\; \pm \;$0.012 & This work; PED fit \\
$T_{\rm eff,estimated}$ (K) & 7211$\; \pm \;$90 & This work; PED fit \\
%$\theta_{\rm LD,estimated}$ (mas) & 0.341$\; \pm \;$0.009 & SED fit \\
\cline{1-3}
$\theta_{\rm UD}$ (mas) & 0.327$\; \pm \;$0.008 & This work \\
$\theta_{\rm LD}$ (mas) & 0.342$\; \pm \;$0.008 & This work \\
$R_{\rm linear}$ ($R_\odot$) &  1.44$\; \pm \;$0.06 & This work \\
$T_{\rm eff}$ (K) & 7193$\; \pm \;$87 & This work \\
$L$ ($L_\odot$) & 5.05$\; \pm \;$0.29  & This work \\
\cline{1-3}
\cline{1-3}
\multicolumn{3}{c}{If the star is contracting towards the ZAMS:} \\
Age (Myr) & $33^{+7}_{-13.2}$ & This work \\
Mass (M$_\odot$) & $1.516^{+0.038}_{-0.024}$ & This work \\
\cline{1-3}
\multicolumn{3}{c}{If the star is expanding from the ZAMS:} \\
Age (Myr) & $90^{+381}_{-50}$ & This work \\
Mass (M$_\odot$) & $1.513^{+0.023}_{-0.024}$ $M_{\odot}$ & This work \\
\enddata
%\tablecomments{}
\end{deluxetable}

\clearpage

%%%%%%%%%%%%%%%%%%%%%%% Figures %%%%%%%%%%%%%%%%%%%%%%%%%%%%%%%

\clearpage

\clearpage

\clearpage

\clearpage

\clearpage

\clearpage

\clearpage

\clearpage

\clearpage

\clearpage

\clearpage

\clearpage

\clearpage

\end{document}